\documentclass{article}   
\usepackage{arxiv}
\usepackage[T1]{fontenc}
\usepackage{xcolor}
\usepackage{graphicx}
\usepackage[natbib=true,backend=biber]{biblatex}
\usepackage{threeparttable}
\usepackage{booktabs}
\usepackage{csquotes}
\usepackage{makecell}
\usepackage{hyperref}
\usepackage{layouts}
\usepackage{pdflscape}
\usepackage{afterpage}
\usepackage{rotating}
\usepackage{float}
\usepackage{geometry}
\usepackage{paralist}
\usepackage{listings}
\usepackage{array}
\usepackage{tikz}
\usepackage{pgfplots}
\usepackage{caption}
\usepackage{subcaption}
\usepackage{pgfplotstable}
\usepackage{multirow}
\usepackage[htt]{hyphenat}
\pgfplotsset{compat=1.17}
\usepackage{enumitem}
\usepackage{pifont}
\usepackage{xstring}
\newcommand{\cmark}{\ding{51}}%
\newcommand{\xmark}{\ding{55}}%

\usepackage{tcolorbox}
\tcbuselibrary{skins}
\tcbset{shield externalize} 
\usetikzlibrary{positioning}
          
\newcommand{\labeltt}[1]{\texttt{\StrSubstitute[0]{#1}{.}{.\allowbreak}}}

\definecolor{tggreen}{HTML}{73d216}
\definecolor{tgred}{HTML}{cc0000}
\definecolor{tgyellow}{HTML}{edd400}
\definecolor{tgviolet}{HTML}{75507b}
\definecolor{tgsilver}{HTML}{d3d7cf}
\definecolor{tgblue}{HTML}{3465a4}

\definecolor{tggreenl}{HTML}{8ae234}
\definecolor{tgredl}{HTML}{ef2929}
\definecolor{tgyellowl}{HTML}{fce94f}
\definecolor{tgvioletl}{HTML}{ad7fa8}
\definecolor{tgsilverl}{HTML}{eeeeec}
\definecolor{tgbluel}{HTML}{729fcf}

\pgfplotsset{
        my boxplot style/.style={
            boxplot,
        },
    }

\newcolumntype{T}{>{\ttfamily}l}

\makeatletter
\newenvironment{btHighlight}[1][]
{\begingroup\tikzset{bt@Highlight@par/.style={#1}}\begin{lrbox}{\@tempboxa}}
{\end{lrbox}\bt@HL@box[bt@Highlight@par]{\@tempboxa}\endgroup}

\newcommand\btHL[1][]{%
  \begin{btHighlight}[#1]\bgroup\aftergroup\bt@HL@endenv%
}
\def\bt@HL@endenv{%
  \end{btHighlight}%
  \egroup
}
\newcommand{\bt@HL@box}[2][]{%
  \tikz[#1]{%
    \pgfpathrectangle{\pgfpoint{1pt}{0pt}}{\pgfpoint{\wd #2}{\ht #2}}%
    \pgfusepath{use as bounding box}%
    \node[anchor=base west, fill=orange!30,outer sep=0pt,inner xsep=1pt, inner ysep=0pt, rounded corners=3pt, minimum height=\ht\strutbox+1pt,#1]{\raisebox{1pt}{\strut}\strut\usebox{#2}};
  }%
}
\makeatother

\newcommand*\circled[2]{\tikz[baseline=(char.base)]{
            \node[shape=circle,draw=#2,inner sep=1pt] (char) {#1};}}

\newcommand{\landscapegeometry}{\newgeometry{vmargin=2cm,hmargin=2cm}%
  \afterpage{\aftergroup\restoregeometry}%
}

\newcommand{\answerspace}{\vspace{0.8\baselineskip plus 0.2\baselineskip minus 0.2\baselineskip}}

\lstdefinestyle{patch}{
    basicstyle=\ttfamily,
    moredelim=**[is][{\btHL[fill=green!30]}]{`}{`},
    moredelim=**[is][{\btHL[fill=red!30]}]{@}{@},
    escapeinside={(*}{*)},
    numbers=left,
    keywordstyle=\ttfamily\bfseries,
    stringstyle=\ttfamily\itshape
}

\newlist{questions}{enumerate}{2}
\setlist[questions,1]{label=RQ\arabic*.,ref=RQ\arabic*}
\setlist[questions,2]{label=\thequestionsi.{\arabic*.},ref=\thequestionsi(\arabic*)}

\newtcolorbox{answerbox}[2][]{
    blanker,
    left=3mm,
    right=3mm,
    borderline west={1.2pt}{0pt}{black},
    title={#2},
    fonttitle=\bfseries,
    coltitle=black,
    #1}

\begin{document}

\title{RunBugRun}
\author{Julian Aron Prenner\\
        \texttt{prenner@inf.unibz.it}\\
         \And
         Romain Robbes\\
         \texttt{romain.robbes@u-bordeaux.fr}
}
\renewcommand{\shorttitle}{RunBugRun -- An Executable Dataset for Automated Program Repair}
\renewcommand{\undertitle}{An Executable Dataset for Automated Program Repair}
\renewcommand{\headeright}{}
\maketitle

 \renewcommand\theadfont{}

\begin{abstract}
Recently, we can notice a transition to data-driven techniques in Automated Program Repair (APR), in particular towards deep neural networks. This entails training on hundreds of thousands or even millions of non-executable code fragments. We would like to bring more attention to an aspect of code often neglected in Neural Program Repair (NPR), namely its execution. Code execution has several significant advantages. It allows for test-based evaluation of candidate fixes and can provide valuable information to aid repair. 
 In this work we present a fully executable dataset of 450,000 small buggy/fixed program pairs originally submitted to programming competition websites written in eight different programming languages. Along with the dataset we provide infrastructure to compile, safely execute and test programs as well as fine-grained bug-type labels. To give a point of reference, we provide basic evaluation results for two baselines, one based on a generate-and-validate approach and one on deep learning.
 With this dataset we follow several goals: we want to lift Neural Program Repair beyond fully static code representations, foster the use of execution-based features and, by including several different languages, counterbalance the predominance of Java in the current landscape of APR datasets and benchmarks.
\end{abstract}


\section{Introduction} \label{intro}

\begin{figure*}[htbp]
\begin{minipage}{0.5\textwidth}
\begin{lstlisting}[language=Go,style=patch,numbers=none,xleftmargin=0pt,xrightmargin=0pt,framesep=0pt]
package main
import "fmt"
func main() {
  // Code for B - ss
  var S string
  fmt.Scanf("%s", &S)
  for i := 1; i < len(S); i++ {
    if (len(S)-i)%2 != 0 {
      continue
    } else if S[0:((len(S)-i)/2)] == \
              S[((len(S)-i)/2):len(S)-i] {
      fmt.Println(len(S) - i)
     (**) `break`
    }
  }
}
\end{lstlisting}
\end{minipage}
\hspace{1cm}
\begin{minipage}{0.4\textwidth}
 \footnotesize

{\renewcommand{\arraystretch}{1.6}%
\begin{tabular}{r|l}
 \textbf{ID} & 780269 \\
 \hline
 \textbf{Language} & Go  \\
 \hline
 \textbf{Split} & Test \\
 \hline
 \textbf{Labels} & \texttt{control\_flow.break.add} \\
 \hline
 \thead[r]{\textbf{Failed}\\\textbf{Tests}} & 
    \tiny 
    {\renewcommand{\arraystretch}{1}%
    \begin{tabular}{c|c|c}
      \textbf{Input} & \textbf{\thead{Expected\\Output}} & \textbf{\thead{Actual\\Output}} \\
      ... & \texttt{6} & \texttt{6\textbackslash n4\textbackslash n2} \\
      ... & ... & ... \\
    \end{tabular}} \\
 \hline
 \thead[r]{\textbf{Passed}\\\textbf{Tests}} & 
    \tiny
    {\renewcommand{\arraystretch}{1}%
    \begin{tabular}{c|c|c}
      \textbf{Input} & \textbf{\thead{Expected\\Output}} & \textbf{\thead{Actual\\Output}} \\
      ... & \texttt{8} & \texttt{8\textbackslash n} \\
      ... & ... & ... \\
    \end{tabular}} \\
  \hline
  \thead[r]{\textbf{Error}\\\textbf{Messages}} & None \\
\end{tabular}}
\end{minipage}
\caption{Example of a bug instance from the presented dataset, along with metadata. Each bug comes with \begin{inparaenum}[a)]
\protect\item the programming language used,
\protect\item the split (i.e., train, valid or test),
\protect\item one or more hierarchical bug labels,
\protect\item a list of failed and passed test cases and
\protect\item one or more error messages in case a runtime error or exception occurred.\end{inparaenum} See Figures~\ref{fig:example-bug-ruby} and \ref{fig:example-bug-java} for more examples.}
\label{fig:example-bug}
\end{figure*}

Not only do software bugs cause costs in the billions~\citep{brittonReversibleDebuggingSoftware, krasnerCostPoorSoftware2020} in the US alone, they also require considerable developer effort~\citep{latozaMaintainingMentalModels2006} and often have turn-around times of several months~\citep{kimHowLongDid2006}. It is thus not surprising that there is great interest in automatically finding and correcting bugs. This is the goal of Automatic Program Repair (APR).

Beginning with GenProg~\citep{forrestGeneticProgrammingApproach2009, gouesGenProgGenericMethod2012}, so-called generated-and-validate (G\&V) approaches dominated APR research over the last decade. Most G\&V-based systems rely on spectrum-based fault localization to first determine a set of suspicious (i.e., likely buggy) code locations upon which the APR tool concentrates its repair efforts.
Spectrum-based fault localization inherently relies on test execution (and consequently also program execution) to determine possible bug locations. In a similar manner, test and program execution is used for evaluation: passing all test cases is a necessary (but not always sufficient) condition for patch correctness.
Thus, in G\&V, execution is a central part of fault localization and correctness evaluation. Sometimes, program execution is even exploited for the actual repairing process. For instance,  DynaMoth~\citep{durieuxDynaMothDynamicCode2016} collects and integrates runtime information into its repairing strategy. Following these lines, prominent APR benchmarks are collections of executable code together with comprehensive test suites (see Section~\ref{sec:related-work} and Table~\ref{tab:apr-benchmarks}).

Recently, more and more APR research builds on deep learning. So much so that this sub-field has been given its own name: Neural Program Repair (NPR).
NPR systems are trained on up to millions of buggy/fixed code fragment pairs. So far, there is a strong focus on static code features, in particular textual features~\citep{chenSequenceRSequencetoSequenceLearning2019, lutellierCoCoNuTCombiningContextaware2020, jiangCURECodeAwareNeural2021}, less commonly, tree or graph representations~\citep{liDLFixContextbasedCode2020, dinellaHOPPITYLEARNINGGRAPH2020}.

Because NPR systems are data hungry and manually collecting and isolating bugs is infeasible on a large scale, copious amounts of bug data are mined from open source code repositories (e.g., GitHub)~\citep{tufanoEmpiricalInvestigationLearning2018}. 

With this dataset we want to combine characteristics of traditional APR benchmarks (executability, test cases) with those of NPR datasets (large size, training data). We motivate our work with: 1) the need for executable code at scale, 2) the need for multi-lingualism, and 3) the need for better curation and insights on performance in large-scale datasets (see Section~\ref{sec:mot}). We summarize these motivations here:

\paragraph{The need for execution at scale.} For NPR datasets, buggy code is usually \enquote{torn} out of context and added to the training set as a mere fragment with a few dozen lines above and below the bug location.
As is usual in machine learning, available data is then divided into three (or more) splits or sets. Usually, \begin{inparaenum}[a)]
\item a training set,
\item a validation set for model parameter tuning and
\item a test set for final performance validation.
\end{inparaenum}
All sets should follow the same data distribution. As the original data is usually comprised of non-executable code fragments, so are validation and test sets. Evaluation must therefore resort to static metrics, that is, the model's result is compared to the ground-truth on a lexical level (e.g. exact match or BLEU). 
However, such static or lexical metrics are problematic because program semantics can be expressed in many different lexical forms (see Figure~\ref{fig:two-fixes}) leading to false negatives. Execution, on the other hand, allows us to run test suites on fix candidates and thus capture correctness on a semantic level.

Many NPR systems are, in addition to the in-distribution test set, also evaluated on established test-based APR benchmarks such as Defects4J~\citep{justDefects4JDatabaseExisting2014} or Bugs.jar~\citep{sahaBugsJarLargeScale2018} allowing for a direct comparison with G\&V systems. However, these benchmarks are not without issues. For one, they are comprised of large Java projects that take long to compile, slowing down evaluation. Because the bugs in such benchmark must often be manually curated (e.g., by adding test cases~\citep{liuCriticalReviewEvaluation2021}) these datasets are relatively small in size, especially in comparison to the large training sets used in NPR. The problem of size is further exacerbated by the fact that many NPR systems are often evaluated only on subsets of said benchmarks (e.g., because they are limited to single-statement, single-line or single hunk bugs)~\citep{zhong2022standup4npr}.
Small benchmarks might also be more susceptible to the problem of benchmark overfitting, a problem reported in previous work~\citep{nodaExperienceReportHow2020a, liuCriticalReviewEvaluation2021}.

\paragraph{The need for multi-lingual APR} We see a strong dominance of the Java programming language in APR. While there seems to be an increasing interest in other languages~\citep{csuvikFixJSDatasetBugfixing2022, widyasariBugsInPyDatabaseExisting2020,gyimesiBugsJSBenchmarkJavaScript2019}, so far, many popular languages are strongly underrepresented in APR (e.g., Go, PHP, Ruby). We advocate for a more multi-lingual orientation of APR. The need for polyglot or language agnostic APR systems is supported by findings in previous work. For instance \citet{camposDiscoveringCommonBugfix2019} found that 44\% of studied open-source GitHub Java projects used more than a single programming language and that 48\% of bug-fix commits modified non-Java source files.

\paragraph{The need for better curation and insights.} One issue with large-scale datasets is that it is difficult to know what they are made of. This has issues in terms of data quality: Are the commits in a large-scale dataset really bug fixes? If they are, are they of an adequate difficulty, and are some type of bug fix over-represented? A second issue is in terms of insights on the performance of APR models, which can direct future improvements. Is a model over-performing on ``simple'' bugs, while under-performing on more difficult bugs? Is a given model performing very well on one category of bugs, but not on others? 

\paragraph{RunBugRun.} In this work, we introduce RunBugRun, a new dataset and benchmark (we consider the dataset's test set a benchmark) for APR, that proposes solutions to the problems pointed out above: 
\begin{itemize}
    \item First, RunBugRun allows execution at scale: as we describe in more detail in Section~\ref{sec:data-col}, 450,000 bugs were collected from short algorithmic programs submitted to competitive programming contests. All of the code is fully compilable and executable. Most importantly, we provide test cases for each bug, along with a Defects4J-like infrastructure to compile, safely execute (inside a sandbox) and test programs.  We have also collected a large number of exception stack traces and error messages that, as recent work showed, can help localize and fix bugs~\citep{yeSelfAPRSelfsupervisedProgram2022}. 
    \item Second, RunBugRun includes programs written in eight different popular programming languages (C, C++, Java, Python, JavaScript, Ruby, Go, PHP) hopefully making this dataset a contribution towards a more polyglot future in APR. Four of the languages are statically typed (C, C++, Java, Go), while the four others (Python, JavaScript, Ruby, PHP) are dynamically typed. Furthermore, the difference in the size of the datasets for each language opens up opportunities to investigate Transfer Learning approaches. 
    \item Third, RunBugRun's 450,000 bugs were obtained from a significantly larger but carefully curated dataset. We ensure that the bugs included are really bugs (with failing test cases) with adequates fixes (where test pass). We also ensure that the bugs are of adequate type (semantic, rather than syntactic bugs) and have an adequate distribution in terms of difficulty. We also address issues such as duplication and flakyness. Moreover, we have labeled most of the bugs into fine-grained categories, allowing to diagnose \enquote{weak spots} in an APR system.
\end{itemize}
In section~\ref{sec:data-col} we explain how we built RunBugRun. The bugs were extracted from Project CodeNet~\citep{puriCodeNetLargeScaleAI2021}, a large dataset of code submissions to programming contest websites. In this section we describe in detail the extensive filtering, pre-processing, execution and labeling steps we undertook.

Section~\ref{sec:the-dataset} presents characteristics of the final dataset including statistical information on programming languages, fix size, bug labels, exceptions and errors as well as data splits. We also present initial result from two APR baselines. One is Cardumen, a Generate-and-Validate approach \cite{cardumen}. The second is an NPR baseline, based on a fine-tuned pre-trained language model, CodeT5 \cite{codet5}. Beyond the raw performance comparison, we also discuss additional findings of our baseline evaluation. We show that there is a strong correlation between repair performance and the number of fixing changes, we highlight strong and weak spot for both baselines, and provide initial evidence that NPR models are able to learn and apply fixes across language boundaries (knowledge transfer). 

From these results, we conclude that there is ample room for improvement for both G\&V and NPR approaches. We discuss some of the way forwards and the challenges ahead in Section~
\ref{sec:discussion}.

Most programs in RunBugRun are short implementations of algorithms to solve specific problems and thus very different from large software projects. We discuss this and other biases and limitations in Section~\ref{sec:limitations}. Finally, we conclude our paper in Section~\ref{sec:conclusion}.

\begin{table*}[htbp] 
	\caption{Overview of automatic program repair benchmarks and datasets. A test-driven evaluation is only possible where tests are provided. Other datasets are mostly used in data-driven approaches (e.g., machine learning)}
	\begin{center}
		\begin{threeparttable}		
			{
				\footnotesize
				\setcellgapes{0.07cm}\makegapedcells
				\begin{tabular}{cccccc}
					\toprule
					\thead{\textbf{Name}} & \thead{\textbf{Type}} & \thead{\textbf{\# Bugs}} & \thead{\textbf{\# Projects}} & \thead{\textbf{Languages}} & \thead{\textbf{Exec.}} \\
					\midrule
					\makecell{Defects4J~\citep{justDefects4JDatabaseExisting2014}} & \makecell{real-world projects} & 835/395\tnote{1} & 17/6\tnote{1} & Java & \cmark \\
					\makecell{Bugs.jar~\citep{sahaBugsJarLargeScale2018}} & \makecell{real-world projects} & 1,158 & 8 & Java & \cmark \\
					\makecell{Bears~\citep{madeiralBearsExtensibleJava2019}} & \makecell{real-world projects} & 251 & 72 & Java & \cmark \\				
					\makecell{QuixBugs~\citep{linQuixBugsMultilingualProgram2017}} & \makecell{algorithm implementations} & 40 & -- & Java, Python & \cmark \\
					\makecell{Codeflaws~\citep{tanCodeflawsProgrammingCompetition2017}} & \makecell{algorithm implementations} & 3902 & -- & C & \cmark \\					
					\makecell{IntroClass~\citep{legouesManyBugsIntroClassBenchmarks2015},\\IntroClassJava~\citep{durieuxIntroClassJavaBenchmark2972016a}} & \makecell{student assignments} & 450/297\tnote{2} & -- & C, Java & \cmark \\
					\makecell{BugSwarm~\citep{tomassiBugSwarmMiningContinuously2019}} & \makecell{failing and passing  CI\tnote{5}\space containers} & 1,940/1,292\tnote{6} & -- & Java, Python & \cmark \\
					\makecell{BugJS~\citep{gyimesiBugsJSBenchmarkJavaScript2019}} & \makecell{real-world  projects} & 453 & 10 & JavaScript & \cmark \\
					\makecell{BugsInPy~\citep{widyasariBugsInPyDatabaseExisting2020}} & \makecell{real-world projects} & 493 & 17 & Python & \cmark \\	
					
					\makecell{Defexts~\citep{bentonDefextsCuratedDataset2019}} & \makecell{real-world\\projects} & 225/301\tnote{7} & 152/170\tnote{7} & Kotlin, Groovy & \cmark \\				
					\midrule
                    \makecell{\textbf{Ours}} & \makecell{algorithm implementations} & \textasciitilde{}450,000 & - & \makecell{C, C++, Java, Python,\\ JavaScript, Ruby, PHP, Go} & \cmark \\									
					
					\midrule
					\makecell{ManySStuBs4J~\citep{karampatsisHowOftenSingleStatement2020}} & \makecell{real world bugs single-line patches (diffs only)} & 153,652 & > 1,000 & Java & \xmark \\
					\makecell{CodRep~\citep{chenCodRepMachineLearning2018}} & \makecell{single-line patches} & 58,069 & -- & Java & \xmark \\
					\makecell{BFP~\citep{tufanoEmpiricalStudyLearning2019}} & \makecell{real-world bugs} & 58,350/65,455\tnote{3} & -- & Java & \xmark \\
					
					\makecell{CoCoNUT~\citep{lutellierCoCoNuTCombiningContextaware2020}} & \makecell{real-world bugs\\diffs and context} & 3,241,966\tnote{4} & -- & \makecell{Java, Python\\JavaScript, C} & \xmark \\				
					
					\makecell{MegaDiff~\citep{monperrusMegadiffDataset600k2021}} & \makecell{real world bugs\\full-file diffs} & 663,029 & -- & Java & \xmark \\
					
					\makecell{FixJS~\citep{csuvikFixJSDatasetBugfixing2022}} & \makecell{real world bugs\\ (mined from GitHub commits)} & 323,907 & -- & JavaScript & \xmark \\
					
					\makecell{TSSB-3M~\citep{richterTSSB3MMiningSingle2022}} & \makecell{real world bugs\\ single-statement} & 3M/9M\tnote{8} & -- & Python & \xmark \\
					
					\bottomrule
					
				\end{tabular}
			}
			\begin{tablenotes}
                \footnotesize
				\item[1] for version 2 and 1, respectively;
				\item[2] for the C and Java versions, respectively;
				\item[3] small and medium versions, respectively;
				\item[4] for Java;
				\item[5] continuous-integration;
				\item[6] for Java and Python, respectively (in September 2021);
				\item[7] for Kotlin and Groovy, respectively;
				\item[8] with and without single-statement/multi-statement \emph{fixes}, respectively;
				
			\end{tablenotes}	
		\end{threeparttable}		
	\end{center}
	\label{tab:apr-benchmarks}	
\end{table*}

\section{Related Work}\label{sec:related-work}
We first cover existing APR benchmarks and NPR datasets, finding a dichotomy between on the one hand small-scale, and executable APR benchmarks, and on the other, large-scale training NPR datasets, that are not executable. Table~\ref{tab:apr-benchmarks} provides an overview of these datasets and benchmarks, contrasting with ours. We then expand on datasets that were extracted from constest submissions, as they are most similar to our work, and compare them to RunBugRun.

\paragraph{APR Benchmarks} In the following we briefly discuss APR benchmarks presented in the recent literature. These benchmarks do not provide any training data and are usually only used for evaluation. All of these benchmarks are executable and come with test cases.

Defects4J~\citep{justDefects4JDatabaseExisting2014} can be considered the most successful APR benchmark and has found widespread usage in the APR literature. It contains 365 (in version 1) and 835 (in version 2) bugs collected from large and prominent Java projects (e.g., Apache Commons libraries) with extensive test suites. Defects4J provides infrastructure to checkout projects in their buggy or fixed state and to run tests.

Bugs.jar~\citep{sahaBugsJarLargeScale2018} and Bears~\citep{madeiralBearsExtensibleJava2019} are two more APR benchmarks of real-world Java bugs. Like Defects4J, Bugs.jar is derived from larger Java projects. The bugs in Bears were found through analyzing failing CI builds of open-source projects on GitHub.

QuixBugs~\citep{linQuixBugsMultilingualProgram2017} is a collection of 40 algorithm implementations (each with a buggy and a fixed version) in Python and Java. Each problem comes with several test cases that can be run with the provided infrastructure. The bugs in QuixBugs are relatively simple, mostly single line and single token bugs. With only 40 bugs it is one of the smallest benchmarks; however it has been used extensively in the APR literature.

With BugsInPy~\citep{widyasariBugsInPyDatabaseExisting2020} and Bugs.js~\citep{gyimesiBugsJSBenchmarkJavaScript2019} there exist two APR benchmarks in Python and JavaScript, respectively. Like Defects4J, both of them consist of bugs in real-world projects and provide testing infrastructure. Similarly, Defexts~\citep{bentonDefextsCuratedDataset2019} is an APR benchmark for Kotlin and Groovy.

BugSwarm~\citep{tomassiBugSwarmMiningContinuously2019} is a collection of pairs of failing and passing CI containers taken from CI-pipelines of open-source Java and Python projects (similar to Bears). It provides infrastructure to download and run containers.

All of the above are mainly evaluation benchmarks. As such, they do not provide training data. The dataset in this work contains, next to a large training set, a test split that can be used as a benchmark. This test split is considerably larger than existing benchmarks (see Table~\ref{tab:apr-benchmarks}).

\paragraph{NPR Datasets}
Unlike benchmarks, the following bug datasets do not contain test cases. Bugs are usually given in the form of non-executable code fragments, that is, either the buggy code line together with a few surrounding lines of code or the method containing the bug. The datasets mentioned are primarily used for training and evaluating NPR models.
ManySStuBs4J is a collection of over 150,000 \enquote{stupid} (e.g., swapped arguments, wrong binary operator) bugs from over 1,000 Java projects. Each bug is classified according to a defect taxonomy.
For their CodeRep machine-learning competition, \citet{chenCodRepMachineLearning2018} collect 58,069 single line bugs from open-source projects. 
\citet{tufanoEmpiricalStudyLearning2019} build a large dataset of buggy/fixed pairs of Java methods, mined from GitHub and used to train a neural network. Similarly, \citet{lutellierCoCoNuTCombiningContextaware2020} mine over three million buggy/fixed code fragment pairs in four different programming languages and use them to train their NPR model. \citet{monperrusMegadiffDataset600k2021} release MegaDiff, a dataset of over 600,000 Java patches whose number of changed lines ranges from 1 to 40.
\citet{csuvikFixJSDatasetBugfixing2022} present FixJS, a dataset of 300,000 pairs of buggy/fixed JavaScript functions from GitHub. Finally, \citet{richterTSSB3MMiningSingle2022} release TSSB, a large dataset of 9 million Python single-statement fixed/buggy Python code fragment pairs. Notably, TSSB-3 is a subset of 3 million pairs where all bugs can be fixed with a single statement change.

\paragraph{Works based on contest submissions}
Most related to this paper is work by \citet{majdCode4BenchMultidimensionalBenchmark2019} who collected a large APR dataset by mining submissions to Codeforces. Our dataset builds on Project CodeNet~\citep{puriCodeNetLargeScaleAI2021}, which, contains submissions from AIZU and AtCoder. There are similarities in particular with regard to data collection. However, we seem to use a much more sophisticated filtering and post-processing regimen. For instance, we ran all submission on all test cases to filter out supposedly correct submissions that fail tests. In their work, only a small subset of submissions is executed. Further, we check all submissions for syntax errors as we are only interested in semantic errors. Thus, all submissions (even buggy ones) in our dataset are compilable. For Python we further convert all Python 2 code to Python 3 code. In some rare cases we even took care of deprecations between minor Python versions (e.g., the \texttt{gcd} function was moved from the \texttt{fractions} package to the \texttt{math} package in Python 3.9). We use a trigram-based metric to make sure there are no cross-set duplicates while they seem to use extact matching. Their dataset seems to provide only a very coarse-grained classification of bugs (i.e., either addition, deletion or modification) we provide fine-grained hierarchical labels for each bug. We further provide performance baselines for a simple NPR  and a G\&V system. Finally, and most importantly, we provide Defects4J-like infrastructure to compile, test and execute all bugs.

\citet{haqueFixEvalExecutionbasedEvaluation2022} compare execution-based and match-based evaluation metrics for NPR. For their study, they collect a dataset of buggy/fixed program pairs from Project CodeNet, which was also used in this work. Again, we find many commonalities in how data was collected. While we are only interested in logical bugs, they also collect submissions with syntax errors. Unlike our dataset, which contains eight languages, the authors only collect submissions in Python and Java. The primary emphasis of \citeauthor{haqueFixEvalExecutionbasedEvaluation2022}'s work seems to be on comparing different evaluation metrics and not on building a dataset or benchmark. Thus, no general infrastructure nor a bug classification seems to be provided.

\citet{tanCodeflawsProgrammingCompetition2017} present Codeflaws, an APR benchmark derived from Codeforce submissions. Their benchmark is monolingual (only C programs) and does not contain training data. With 7,436 bugs it is also considerably smaller than RunBugRun (over 30,000 bugs in the test set). Notably, Codeflaws categorizes each defect into a taxonomy of 40 different defect classes.

\section{Motivation}\label{sec:mot}
In this section we explain in detail why we see the need for a new dataset for APR. This is based on three main motivations: 1) the need for a large-scale execution dataset and its implications (Section 3.1), 2) the need for a multi-lingual APR dataset (Section 3.2), and 3) the need for curation and labelling (Section 3.3).

\subsection{APR needs a large-scale \emph{and} executable dataset}

APR is comprised of two sub-communities, that different data requirements have been pushing apart. On the one hand, classical G\&V approaches rely on executable datasets; these executable datasets are hard to scale, due to manual evaluations and long execution times, and are at risk of benchmark overfitting. On the other hand, NPR approaches use large-scale datasets for training. These are non-executable, preventing the model to learn valuable information from program execution, and forcing it to rely on lexical information for evaluation (with the exception of executable benchmarks); moreover, the large scale and the lack of execution causes some data quality issues. 

Thus, we argue that a dataset that is made of a large number of small, executable, and well-tested programs is a good middle ground. Such a dataset can scale, as the run-time evaluations are not too costly. Scale allows for more diversity in the types of bugs, reduces the risk of benchmark overfitting, and offers an alternative to manual evaluation. Finally, the availability of execution data at training time offers some potential for NPR to learn from new sources of information. We detail each of these points in the following paragraphs.

\paragraph{Moving NPR and G\&V closer}
As previously mentioned, most established APR benchmark have been created at a time when G\&V approaches were at the center of research interest. Consequently, these benchmarks fulfilled the needs G\&V systems: they are fully executable, in order to allow spectrum-based fault localization as well as running tests for evaluation. Collecting and curating large number of high-quality bugs is very tedious. It is thus not surprising that these benchmarks are rather small by today's standards.

Increasingly popular neural-based methods (NPR), on the other hand, are trained on large amounts of data. As is customary in machine learning, a certain proportion of the data (usually between 5-15\%) is set aside and used for validating (e.g., tuning hyper-parameters) and evaluating the model. NPR borrows many techniques from neural natural language processing (NLP), frequently relying on the same model architectures and tokenization algorithms. In many cases, NPR also follows NLP evaluation practices by employing static metrics such as BLEU, ROUGE or $n$-gram similarity. However, unlike natural language, code is executable. By providing a relatively large fully executable dataset, NPR and traditional APR methods can share a common evaluation regimen, facilitating direct performance comparisons.

\paragraph{Benchmark overfitting issues} Because existing APR benchmarks come with no training data, bugs in the benchmark might be used for tuning and optimizing APR systems (e.g., by focusing on bug patterns appearing in these benchmarks). This may cause biases towards specific benchmarks. Previous work found significant performance gaps between different benchmarks as well as between real-world bugs and bugs in benchmarks (benchmark overfitting).
For instance, \citet{nodaExperienceReportHow2020a} found Elixir~\citep{saha2017elixir} to underperform in an industrial setting, relative to its performance on common benchmarks.
Similarly, \citet{liuCriticalReviewEvaluation2021} raised concerns about evaluating an APR tool's performance \enquote{in-the-lab} versus \enquote{in-the-wild}. Finally, \citet{durieux2019empirical} in an extensive review found that many APR tools overfit to the Defects4J benchmark. 
Providing fully executable train and validation sets allows G\&V approaches to follow a more strict data separation regimen (as is customary in NPR), possibly mitigating benchmark overfitting and other biases.

A second issue with overfitting is that, when using a small evaluation dataset (with dozens to hundreds of defects), too much importance might be given to the performance of a few datapoints, which might affect the reliability of conclusions. Of note, this also affects NPR systems when they are evaluated on standard APR benchmarks. A larger evaluation dataset would be more robust to this, if the dataset can allow for evaluation at scale, which is the focus of the next points.

\begin{figure}[htbp]
    \centering
\begin{tabular}{c}    
\begin{lstlisting}[language=Ruby,style=patch]
n = gets.chomp.to_i
n.times do |i|
(*\hspace{1em}*) line = gets.chomp
(*\hspace{1.2em}*) @line.gsub(/Hoshino/, "Hoshina")@
(*\circled{1}{black}*) `line.gsub!(/Hoshino/, "Hoshina")`
(*\circled{2}{black}*) `line = line.gsub(/Hoshino/, "Hoshina")`
(*\hspace{1em}*) puts line
 end 
\end{lstlisting}
\end{tabular}
\caption{Two alternative patches for a Ruby bug in our dataset. The result of line 5 remains unused. Patch 1 (line 5) uses a method variant that modifies \texttt{line} in-place; patch 2 (line 6) reassigns the result to \texttt{line}, leading to the same result. This shows that determining patch correctness statically (i.e., without execution) often results in false negatives.}
\label{fig:two-fixes}
\end{figure}

\paragraph{Test-based evaluation improves over textual evaluations}
As explained previously, NPR is largely following neural NLP techniques, including evaluation practices. However, unlike natural language, code can be executed. In traditional APR, execution is used for fault localization and test-based evaluation. 
To see why tests are useful for patch correctness evaluation, consider the case where, in order to fix a bug, the expression \texttt{x} must be transformed into the expression \texttt{x/2}. However, an APR tool transforms the buggy expression \texttt{x} into the \emph{logically} but \emph{not lexically} equivalent expression \texttt{x * 0.5}. As both solutions lead to the same program output, a test case will treat both solutions equally. However, lexical metrics such as exact match cannot easily capture such semantic equivalences and \texttt{x * 0.5} would be considered an incorrect fix.
Figure~\ref{fig:two-fixes} shows another case, where a lexical comparison cannot correctly determine patch correctness. The bug can be fixed in two different ways: either we reassign the result from the \texttt{gsub} method call to the \texttt{line} variable or we use the in-place variant of the same method (indicated by the exclamation mark).

In a recent study, \citet{haqueFixEvalExecutionbasedEvaluation2022} compared static match-based metrics with test-based (or execution-based) metrics for evaluating NPR models. The former include metrics such as exact match, BLEU/CodeBLEU~\citep{codebleu}, compilability rate or AST subtree overlap and operate on a lexical or syntactical level, while the latter rely on test execution (e.g., average number of passing test cases). The authors find that execution-based metrics
\enquote{\textelp{} are better performance indicators for functional program correctness and evaluate models better than match-based metrics}. In particular, the authors show that static (i.e., lexical or syntactic) metrics can be deceiving. A dummy naive copy baseline (i.e., output the buggy code as potential fix) can obtain higher values for the BLEU, CodeBLEU, AST subtree overlap and compilation rate metrics than a CodeT5~\citep{codet5} or PLBART~\citep{plbart} model fine-tuned on the repair task, despite this baseline never outputting any correct fix (by design). On the other hand, execution-based metrics clearly reveal the bad performance of the copy baseline. Our dataset contains executable programs and test cases, allowing the use of execution-based metrics.

\paragraph{Test-based evaluation should be made more efficient}
The overhead of running large test suites is one of the major drawbacks of test-based evaluation. For instance, \citet{chenContractbasedProgramRepair2017} report that patch validation takes up 92.8\% of their APR tool's running time. While some other APR tools try to cut down validation time (e.g., by skipping irrelevant test cases~\citep{yuanARJAAutomatedRepair2020} or compiling multiple patch candidates in one go~\citep{huaSketchFixToolAutomated2018}), this involves a substantial engineering effort that must be repeated for every APR tool.
Moreover, many of the current datasets containing tests are comprised of bugs in large projects. In order to run the tests, the entire project codebase must be compiled and possibly a large number of project dependencies must be installed. This further slows down test execution besides consuming a lot of hard disk space.
One solution to this problem is the use of small programs with no or few external library dependencies. A quicker compile-run-test cycle allows up to hundred program execution per second (on machines with multiple CPUs) which in turn makes it possible to evaluate a much larger number of bugs. RunBugRun provides infrastructure to safely (in a sandbox) execute and evaluate the programs in the dataset. Because programs are relatively small and multiple of them can be compiled and executed in parallel, evaluation is several magnitudes faster than is the case with Defects4J, where  heavy test suites and build systems slow down compilation, execution and testing.

\paragraph{Scaling issues with manual evaluations}
Passing all test cases is no guarantee for a patch to be fully correct~\citep{smithCureWorseDisease2015, qiAnalysisPatchPlausibility2015}. Unfortunately, this means that developers need to manually inspect each patch generated by their APR tool and assess it for correctness. This is a painstaking task that requires considerable time and effort. Moreover, since manual evaluations are often subjective and subject to errors \cite{ye2021automated, le2019reliability}, it is common practice to need multiple annotators to estimates disagreements, which multiplies the effort.
With larger test sets and benchmarks, this manual task becomes infeasible. 

We argue that large-scale executable datasets with strong test suites constitute a reasonable trade-off. A strong test suite with a high coverage makes it unlikely that incorrectly patched programs pass all tests. While this is not a perfect solution, manual evaluations can contain errors as well. Given that the programs included in RunBugRun are relatively small, having stronger test suites that run in a reasonable time is possible. It is also possible to strengthen test suites by exploiting the fact that problems have multiple solutions: we can generate additional test cases with relatively little effort by running multiple (correct) programs solving the same problem on random input. If they all output the same result, it can be assumed with very high confidence to be the correct answer and used as a test case (see Section~\ref{sec:data-col}).

Finally, switching to an automated evaluation would not only lift the burden of manual assessment off developer's shoulders but also allow evaluating APR systems on a much larger number of bugs. This increase in scale would thus lead to more reliable performance metrics by reducing the issues of overfitting and excessive reliance on individual data points.

\paragraph{Real-world vs. artificial bugs}
An alternative to scale program repair dataset is to \emph{seed} artificial bugs (e.g., by generating mutants). While it is relatively easy to generate large numbers of such artificial bugs, collecting real-world bugs usually involves a substantial mining, filtering and pre-processing effort (as described below). However, real-world bugs are more varied than artificial ones, since the artificial bugs usually come from a set of hand-crafted mutations. Indeed, existing work indicates that artificial bugs are not a full-fledged replacement for real-world bugs~\citep{richterCanWeLearn2022, allamanisSelfSupervisedBugDetection2021a,bug-detector-dist-shift}, which is why RunBugRun focuses on real-world bugs. 

That said, RunBugRun might serve as a foundation for further research in this direction, by, say, extending the training set with seeded bugs or comparing performance on the test set with and without additional bug seeding.

\paragraph{Potential of executable code}
Last but certainly not least, we believe that integrating execution-based features into NPR models has great potential. Previous work already reported promising results. For instance, early work by \citet{yeSelfAPRSelfsupervisedProgram2022} shows that error messages and stack traces obtained through execution are a valuable information source that can improve repair accuracy. Similarly, \citet{drainDeepDebugFixingPython2021} report a performance boost when including Python stack traces as training data into their NPR system. Of course, such information can only be obtained through code execution. Recent work by \citet{yeNeuralProgramRepair2022} shows that integrating feedback from test execution \emph{during training} can improve repair performance. They combine a NMT-based NPR model with a discriminator component that, for a small number of executable bugs with tests, compiles patch candidates and runs them against a test suite; whether the candidate is compilable and how many tests it passes is then fed back into the overall model and used a learning signal. Using execution feedback, the authors were able to increase overall performance by 7\%. However, because executable bugs are hard to come by, such feedback could only be obtained for 123 bugs. By providing several hundred thousand of such executable bugs, our dataset opens up new opportunities for this line of research.
Leveraging execution feedback also showed promising results in the field of program synthesis~\citep{coderl}.

Execution is a rich source of useful information. For instance, execution traces can give insight into program state changes. We envision a new generation of APR tools that can exploit runtime information for better bug localization and repair. 
Finally, executable ground-truth programs can be used to  generate an arbitrary amount of artificial test cases.

\subsection{APR needs a multilingual dataset to overcome language bias}

In recent years, we have seen a breakthrough in natural language processing (NLP) and understanding (NLU), largely owed to deep learning. However, researchers have repeatedly indicated that there exists a strong bias towards the English language, not only in dataset but also benchmarks (e.g., GLUE)~\citep{benderDangersStochasticParrots2021}.

Does such a bias exist in APR? Of the 16 most important datasets published in recent years, 11 include Java, five Python and three JavaScript, while other less popular but still very common languages like Ruby, PHP or Go do not appear at all (Table \ref{tab:apr-benchmarks}).
We would like to open the field to a larger number of programming languages, overcome its current Java centricity and foster the development of more polyglot or even language-agnostic APR systems. This would  greatly raise the applicability of APR because large amounts of code is written in languages other than, say, Java, Python or C.
With software development becoming increasingly diverse in terms of programming languages, bugs might involve several different programming languages~\citep{zhong2015empirical}. In fact,  a recent study found this to be surprisingly common ~\citep{camposDiscoveringCommonBugfix2019}.
By putting more emphasis on polyglot repair, future APR systems might be more successful at fixing such multi-language bugs.

In NLP, data availability is distributed very unevenly over different languages~\citep{joshiStateFateLinguistic2020}. 
Neglect of languages with low data availability, in particular of minority languages, can be a threat to their survival and decrease linguistic diversity~\citep{ruder2020beyondenglish}.

A similar pattern can also be observed for code, where there is more code written in older (e.g., C++) or more popular languages (e.g., Python) than in niche languages such as Ruby.
Focusing technological progress on only one or a few languages may cause developers to select programming languages based on the quality and availability of tooling and not because of the language's fitness for a particular task. This might cement and reinforce the use of established programming languages and stymie the adoption of new or more suitable ones.
With eight different programming languages we hope that our dataset serves as a basis for a future APR research that crosses language boundaries.

\subsection{APR needs a curated and labelled dataset---and execution helps}  

\paragraph{Mining patches from commits}
\begin{figure}[htbp]
    \centering
\begin{tabular}{c}    
\begin{lstlisting}[language=python,style=patch,breaklines=true, showstringspaces=false]
# General information about the project.   
project = u'GitPython'

copyright = u'Copyright (C) 2008, 2009 Michael Trier and contributors, 2010(*\btHL[fill=green!30]-2015*) Sebastian Thiel'

# The version info for the project you're documenting, acts as replacement for
# |version| and |release|, also used in various other places throughout the
\end{lstlisting}
\end{tabular}
\caption{A fix from the TSSB-3M dataset~\citep{richterTSSB3MMiningSingle2022}, with three million instances, currently one of the largest APR datasets available. The dataset contains non-executable code fragments. Although this dataset instance is marked with \enquote{likely bug}, the change (updating of the copyright year) is unlikely to resolve a functional bug.}
\label{fig:false-positive}
\end{figure}

Bugs and corresponding human-written patches are often mined from code repositories. Unfortunately, repository commits are often entangled~\citep{herzigImpactTangledCode2013}, meaning that a single commit addresses multiple concerns and thus may contain changes not related to the actual bug fix. While simple techniques using keyword-matching on commit messages have shown high precision for extracting bug-fixing commits, such methods also exhibit very low recall~\citep{babiiMiningSoftwareRepositories2021}, making it harder to gather large amounts of high-quality patches for today's data-hungry models. When mining large numbers of commits, false positives are hardly avoidable. This means that most commit-mined APR datasets contain a (small) portion of instances that are actually not bugs or bug-fixes. Such a case is shown in Figure~\ref{fig:false-positive}, taken from TSSB-3M~\citep{richterTSSB3MMiningSingle2022}, where updating the copyright year was mistakenly identified as bug-fixing change. \citet{lutellierCoCoNuTCombiningContextaware2020} found, by manually inspecting a small sample of their commit-mined APR dataset, that 7\% of dataset instances in the sample were not actual bug fixes. A similar check was done by \citet{tufanoEmpiricalStudyLearning2019} who found only 2.4\% of false positives in a sample of 384 instances.

Having a dataset that is both large-scale and executable helps in this regard: through execution we can warrant that the change actually fixes a bug: the buggy version should have failing test cases, while the fixed one only passing test cases. We note that this only a starting point, as additional steps toward curation are necessary.

One such step is the need for \emph{labelling} data: a very large collection of buggy/fixed program pairs, by itself, is of limited usefulness if we do not know what are the types of bugs that are in it. If the bugs are overly simple or are not diverse enough, the performance of a model on such a dataset might be misleading. To ensure a minimum amount of variety in terms of bug type and visibility, this type of information needs to be inferred somehow. Moreover, having access to fine-grained performance on multiple types of bugs can provide some insight on model performance and how to improve it down the lines (e.g., a model that perform less well on bugs involving operators might be improved in a different way than one that underperforms on fixing function calls). Some of the possible ways to label the data include the size of the bug fix, the type of repair that was made, or, if execution data is available, the type of error that was triggered by the faulty program.

\section{Methodology}\label{sec:data-col}

Our dataset is based on Project CodeNet~\cite{puriCodeNetLargeScaleAI2021}, a large, multi-lingual data dump of programming contest submissions.
However, this data is rather \enquote{raw}. For this reason, it had to be heavily pre-processed and filtered, with two major objectives: 1) ensuring that all the programs in the datasets can be (safely) executed, and that they have adequate test coverage, 2) ensuring that the data is properly curated and labelled to constitute a dataset of sufficient quality and diversity in a set of metrics of interest, and 3) providing a benchmark with an adequate difficulty (not too easy, but also not too difficult), and initial baselines. We present all the steps we took in chronological order, rather than grouped by objectives, as this seems to be the most natural order.

\paragraph{Project CodeNet} Our dataset is derived from Project CodeNet~\citep{puriCodeNetLargeScaleAI2021}, a large collection of programming contest submissions mined from two online judge websites (AIZU and AtCoder). These websites publish a list of coding problems comprised of a problem description and example input and output. Users can then submit their solution programs for a particular problem. Submitted programs are automatically evaluated by the website's judging system using a non-public set of test cases. Project CodeNet contains over 13 million submitted programs for over 4,000 problems. Each submission is, among other information, associated with an anonymized user id, a timestamp and a judgement (i.e., whether the submission correctly and completely solves the problem or not). If the submission passes all tests it is \emph{accepted}, otherwise it is \emph{rejected}.  

\paragraph{Filtering for Language (curation)}
As a first step, we remove all submissions written in a programming language that is not one of C, C++, Java, Python, Ruby, Go or PHP. Our selection of programming languages follows CodeSearchNet~\citep{husain2019codesearchnet}, a popular machine learning dataset for code that contains code in Java, Python, Ruby, PHP and Go. We also add C and C++ because a large number of submissions are written in these languages.

\paragraph{Selecting likely buggy/fixed program pairs (curation)}

To extract bugs from Project CodeNet we first group the remaining submissions by problem and user id. Then, for each possible pair of $(s_1, s_2)$ of a rejected submission $s_1$ and an accepted submission $s_2$ belonging to the same group (i.e., $s_1$ and $s_2$ belong to the same problem and user) we calculate the number of different tokens between them. This difference is calculated by first tokenizing both submissions into a sequence of tokens. Then we use a LCS-based diffing algorithm to compute the token-level difference. If the difference is less than six tokens we consider $s_1$ a buggy submission and $s_2$ a corresponding fixed submission and add them to our dataset.
A similar approach was used in previous work~\citep{haqueFixEvalExecutionbasedEvaluation2022, code4bench}. We use a relatively low threshold for the maximum token difference such that our mining process yields bugs that can be fixed with few changes. This choice reflects the fact that current state-of-the-art APR system still struggle with bugs that require multiple change actions~\citep{yangWhereWereRepair2021, bennett2022some}. Once APR systems are able to correctly fix, say, more than 90\% of bugs this threshold could be raised to create a more difficult dataset.

\paragraph{Mining test cases (ensuring executability)}
All code submissions read an input string from standard input and output the answer to standard output. A test case is thus a string pair $(i, o)$, where $i$ is the input and $o$ the expected output. A test asserts that, for a given program $p$, $p(i) == o$.
Unfortunately, Project CodeNet does not contain the test cases used by the online judge websites to accept or reject submissions. However, each problem description (written in HTML) contains 1-6 exemplary  input/output pairs. Project CodeNet comes with a script to extract the first example pair from a problem's description file. To extract all example input/output pairs we implement our own extraction scripts with the help of which we extract 9,082 example pairs which we use as test cases. Unfortunately, this number is fairly low (roughly two test cases per problem). To further increase the number of tests we add additional test cases found in CodeContests\footnote{\url{https://github.com/deepmind/code_contests}}, the dataset used to train AlphaCode~\citep{alphacode}. CodeContests stores data as compressed protocol buffers. Once decompressed and decoded, they could be easily extracted as CodeContests and Project CodeNet use the same problem ids. This adds 331,834 more test cases. Finally, for problems with low test count we use a semi-automatic process to generate further test cases (see below). Figure \ref{fig:tests-per-problem} in the appendix shows the distribution of the number of test cases.

\paragraph{Filtering out submissions with syntax error (curation)}
As we are targeting \emph{semantic} program repair, our goal is to compile a dataset of semantic (or logical) bugs that are syntactically valid.
Therefore, we filter out all pairs $(s_1, s_2)$ that contain syntax errors (in either $s1$ or $s2$). For languages that require a compilation step (C, C++, Java, Go) we compile each submission and remove $(s_1, s_2)$ if compilation fails. Scripting languages are checked for syntax errors using special \enquote{syntax-check only} modes supported by all interpreters in question.
We would like to note that a similar approach (i.e., only keep submissions with syntax errors) could be applied to obtain a dataset of \emph{syntactic} bugs. This is, however, not within the scope of this work.

\paragraph{Pre-processing (ensuring executability) }
Most submissions do not require pre-processing. However, we found that a large number of Python submissions where written for Python 2. We use the \texttt{2to3} utility to automatically convert all Python 2 submissions to Python 3 compatible code (e.g., replacing \texttt{raw\_input()} with \texttt{input()}). In order to ensure compatibility with recent versions of Python we take account of a change introduced in Python version 3.9 in which the \texttt{gcd} function was moved from the \texttt{fractions} to the \texttt{math} module.

\paragraph{Execution (ensuring executability)}
We execute all remaining submission on all test cases belonging to the corresponding problem. We record the result of the execution, in particular, the program's output and error messages in case of error. All executions are done within a \emph{sandbox} that protects the host system from possible malicious code. Our sandbox is based on \texttt{bwrap}\footnote{\url{https://github.com/containers/bubblewrap}} which is also used by Flatpak, a package distribution and management system for Linux. We configure \texttt{bwrap} in such a way that submissions cannot
\begin{inparaenum}[a)]
\item connect to a network or the Internet,
\item write or read files from a user's home directory and
\item read only a number of pre-defined system files that are required for program execution (e.g., compiler and interpreter executables, standard libraries etc.)
\end{inparaenum}.
The following versions of compilers and interpreters are used:
\begin{inparaenum}[a)]
\item GCC 11 for C and C++,
\item CPython 3.9 for Python,
\item CRuby 3.0 for Ruby,
\item OpenJDK 17 for Java,
\item PHP 8.1.2,
\item Go 1.18.1 and
\item NodeJS 12.22
\end{inparaenum}.
If submission execution does not complete within 3 seconds it is aborted and marked as \enquote{timed out}. We also restrict the amount of available memory to 512 megabytes by either limiting a process' virtual memory through the operating system or by setting an appropriate VM/interpreter parameter (JVM, NodeJS).
In total, we recorded over 130 million executions; this took several weeks.

\paragraph{Submission and test case filtering after execution (curation)}
Executions serve as basis for further filtering. We remove any pair $(s_1, s_2)$ if
\begin{inparaenum}[a)]
\item $s_1$ passes all test cases or times out on any test case or
\item $s_2$ fails any test case or times out on any test case.
\end{inparaenum} Moreover, we filter out test cases that 
\begin{inparaenum}[a)]
\item have an overall pass rate below 45\% or
\item have a timeout rate above 12\% (i.e., more than 12\% of submissions time out if fed the test case's input).
\end{inparaenum}
When choosing these thresholds we tried to find a good balance between keeping as many tests as possible and filtering out tests with low pass or high timeout rates. We found 45\% and 12\% to provide good trade-offs. The distributions of the pass and timeout rates (Figure~\ref{fig:test-rates} in the appendix) may further elucidate the choice of these thresholds.

\paragraph{Generating test cases using execution (ensuring executability)}\label{sec:testgen}
Even after adding all test cases from the description files and from AlphaCode's CodeContests some problems remain with no or very few test cases. For these problems we generate additional test cases in a semi-automatic way. For each problem, we write a script that generates random inputs (adhering to the problem specification). We than sample 10-15 (depending on the problem) random \emph{correct} (according to the online judge) submissions to which we feed the generated input. If all sampled and correct submissions yield the same output we consider the output to be correct and add it, together with the generated input, to our list of test cases. This way, we generate over 600 additional test cases for 111 problems.

\paragraph{Labeling (curation)}
We automatically label filtered bugs, that is buggy/fixed submission pairs $(s_1, s_2)$, with one or more labels. Our labeling system is rule based. We parse both $s_1$ and $s_2$ using the TreeSitter parsing library\footnote{\url{https://tree-sitter.github.io/tree-sitter/}} (which supports all languages in our dataset). We then find all nodes in the parse trees affected by changes between  $s_1$ and $s_2$ (i.e., a specific node was added, removed or replaced). This is done by first tokenizing both $s_1$ and $s_2$ and then calculating a LCS-based token-level diff. For each changed, inserted or deleted token we then lookup the corresponding node in the parse tree. Next, the changed nodes are matched against a set of manually written rules. Each rule, if matched, marks the bug with a corresponding label. For instance, say, a break statements was added to a loop in $s_2$ to fix the bug in $s_1$. The set of changed nodes will be a singleton set containing only a node of type \texttt{statement\_break}. A matching rule is found: it matches because \begin{inparaenum}[a)]
\item a node was added and
\item the node is of type \texttt{statement\_break}.
\end{inparaenum}
Consequently, the bug is labeled with \texttt{control\_flow.break.add}.
Labels are hierarchical, that is, are grouped with common prefixes. For example, all control flow related labels (e.g., breaking out of a loop, skipping an iteration, adding or removing a if-else branch) share a common prefix (see Figure~\ref{fig:labels-diag} in the appendix for a diagrammatic overview of the label hierarchy).
Since parse trees can differ between languages, the same rule must often be implemented for different languages. Our \enquote{change taxonomy} contains 136 different leaf labels and 16 stem labels (i.e., prefixes). 
In some cases, changes are of \enquote{singular kind} and not matched by any rule. This is the case for roughly 9\% of submissions. On average, each bug is labeled with 2.3 labels; 25\% of data instances have one label, 26.8\% two.

\paragraph{Excluding flaky submissions (curation)}

Previous work has found flaky tests to negatively affect bug localization and to lower repair success~\citep{flaky}.
We check all submissions in the test and validation sets for flakiness, that is, whether they randomly change output upon repeated execution. 
To this end, we run all fixed submissions belonging to these sets several times (at least five times). If any run fails to pass all test cases it is considered flaky and consequently removed from the dataset. We find that the main source of flakiness are memory errors in C and C++ (e.g., double free or invalid memory access). In a single Python submission, flakiness came from improper use of pseudo random numbers (without seeding).

\paragraph{Data splitting (benchmarking)}
In order to allow an apples-to-apples comparisons we split the data into pre-defined train, test and validation sets that future users of the dataset ought to use. Most submissions are written in C++ and Python. However, to counter balance the overweight of these languages we make sure that in the test set the language distribution is more balanced. In the test set, the ratio between the most and least dominant languages (C++ and Python versus Go, JavaScript and PHP) is roughly 10x, while this ratio is over 100x  in the overall data. As mentioned previously, each submission is associated with an anonymized user id. We split the data in such a way that multiple dataset instances belonging to the same problem and user are never split across different sets.

We decide to split \emph{across} the problem dimension. That is to say, submissions solving a given problem are divided among the three sets. Because these submissions solve the same problem they might bear a certain similarity (below the deduplication threshold). We chose this setting as one of the central ideas of APR is to \enquote{transplant} correct code into buggy code. In a way, we test an APR system's ability to learn and pick up fix patterns from the training set and apply them to the test set. This is to some extent analogous to the plastic surgery hypothesis~\citep{plasticsurgery} in that the fix ingredients can be learned from programs in the training set.

An alternative that we considered would be to emphasize generalization further, by having only unseen problems in the test set. However we thought this setting would have been too challenging for the current state of APR/NPR. We plan to realise a follow-up dataset using this alternative format in the future, on performance on RunBugRun starts to saturate.

\paragraph{Deduplication (curation)}
While we have submission to the same problems in the training and test sets, we still want to avoid data leakage. To do so, we take the following deduplication measures: \begin{inparaenum}[a)]
\item we remove all exact duplicates (ignoring whitespace) from the entire dataset and
\item we remove cross-set near duplicates between the training and the test set as well as between the training and the validation set.
\end{inparaenum}
We use trigram similarity to efficiently find near duplicates on hundred of thousands of submissions in eight different programming languages.

\paragraph{Baselines (benchmarking)}
In order to give future users of the dataset a reference point in terms of expected performance we provide two simple baselines: a NPR system based on CodeT5~\citep{codet5} as well as a G\&V baseline. As G\&V baseline we choose Cardumen~\citep{cardumen} because it is one of the most successful G\&V systems on QuixBugs~\citep{YE2021110825}, a benchmark that is similar to our dataset in terms of code type and bugs (i.e., simple bugs in algorithm implementations).
Please refer to Section~\ref{sec:baseline-eval} on methodological details of our evaluation.

\section{Results \& Discussion}\label{sec:the-dataset}

Our dataset contains over 450,000 bugs, that is buggy/fixed program pairs derived from submissions to programming contest websites. Each bug is associated with \begin{inparaenum}[i)]
\item a unique bug identifier
\item a language identifier (e.g., \texttt{java}),
\item one or more optional labels (9\% of bugs do not not fall into an existing label),
\item a set of test cases in the form of input/output pairs,
\item zero or more error messages raised by the buggy program
\end{inparaenum} (see Figures~\ref{fig:example-bug}, \ref{fig:example-bug-ruby} and \ref{fig:example-bug-java}).

We investigate the main research questions:
\begin{questions}
\item \textit{Dataset Statistics:} Here we want to find out what the most important key figures of the dataset are. To put it bluntly, we want to know \enquote{what's inside}.
In particular, we try to answer the following: 
\begin{questions}
\item How are languages distributed?
\item What is the distribution of the number of changes?
\item What are the most common bug labels?
\item What are the most common errors and exceptions that occurred during test executions.
\end{questions}

\item \textit{Baselines:}
The second research question is concerned with baseline performance. The goal is to provide future users of our dataset with a yardstick by which their tools' performance can be measured. 
Finally, for our NPR baseline, we want to find out if fixes can be learned in one language and applied to another (i.e., transferred).
This is particularly important in view of the uneven language distribution where there are large amounts of data for some languages (e.g., C++ or Python) and very small amounts for others (e.g., PHP or Go). In more detail, we ask:
\begin{questions}
\item How well do the selected baselines perform on the dataset? How does the performance vary across languages and number of changes/hunks?
\item How does performance vary across labels and what are the strengths and weaknesses of the NPR and G\&V baseline, respectively?
\item How well do bugs \textit{transfer} from one language to another, i.e., how well can we learn bugs in one language and fix them in another one?
\item How does the number of generated candidates affect performance?
\end{questions}
\end{questions}

\subsection{Dataset Statistics (RQ1)}

\subsubsection{Language Distribution (RQ1.1)}

\begin{figure*}[htbp]
    \centering
	\includegraphics[width=\linewidth]{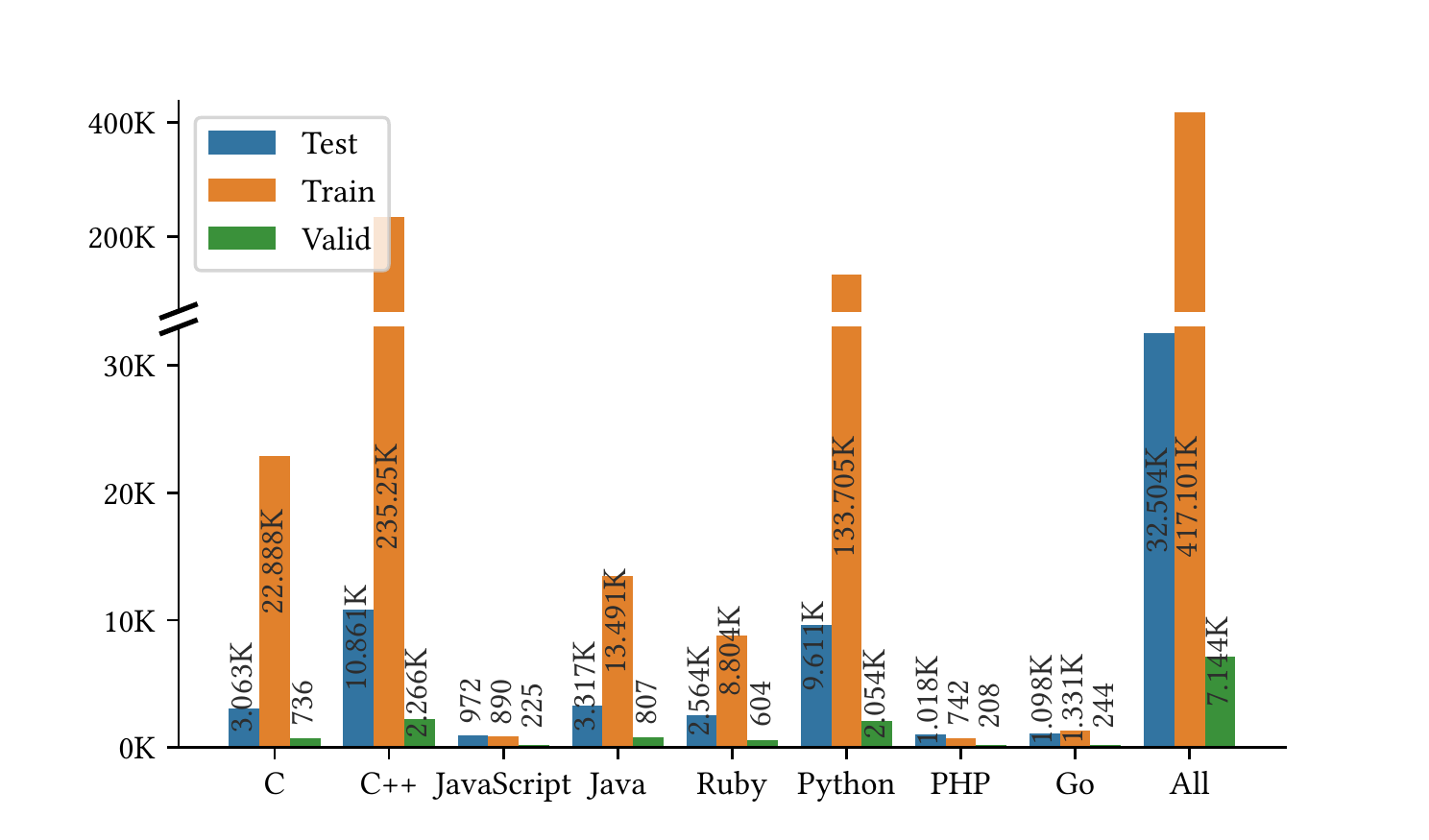}
    \caption{Language distribution for the training, test and validation set, respectively.}
    \label{fig:lang-dist}
\end{figure*}

Our dataset contains eight different languages (C++, C, Python, Java, Ruby, PHP, JavaScript, Go). Unfortunately, there is a disproportionate number of submissions written in C/C++ and Python and very few submissions for PHP, JavaScript and Go. To counteract this imbalance, the test set contains, relatively speaking, a larger percentage of underrepresented languages (at least 1,000 instances). As we show in a later section, modern NPR models are able to learn fix patterns in one language and apply them to other languages (RQ2). Thus, despite this imbalance, we think that our dataset is an important contribution to future polyglot APR systems. Figure~\ref{fig:lang-dist} shows the language distribution in the training, test and validation sets.
C++ and Python are the most common languages, with over 100,000 instances in the training set, followed by C with over 23,000. In terms of size, Java and Ruby are in the mid-range with over 8,000 and 13,000 instances in the training set, respectively. PHP and JavaScript both have training sets of slightly below 1,000 instances, Go slightly above.

\vspace{0.8\baselineskip plus 0.2\baselineskip minus 0.2\baselineskip}
\begin{answerbox}{Answer to RQ1.1}
The language distribution is heavily skewed towards C/C++ and Python. JavaScript, Go and PHP are the least frequent languages. We compensate this imbalance by increasing the proportion of low-frequency languages in the test set such that the ratio between the most and least frequent language is roughly 10x (this ratio is approximately 125 is the overall data).
\end{answerbox}

\subsubsection{Number of Changes (RQ1.2)}

\begin{figure*}[htbp]
    \centering
	\includegraphics[width=\linewidth]{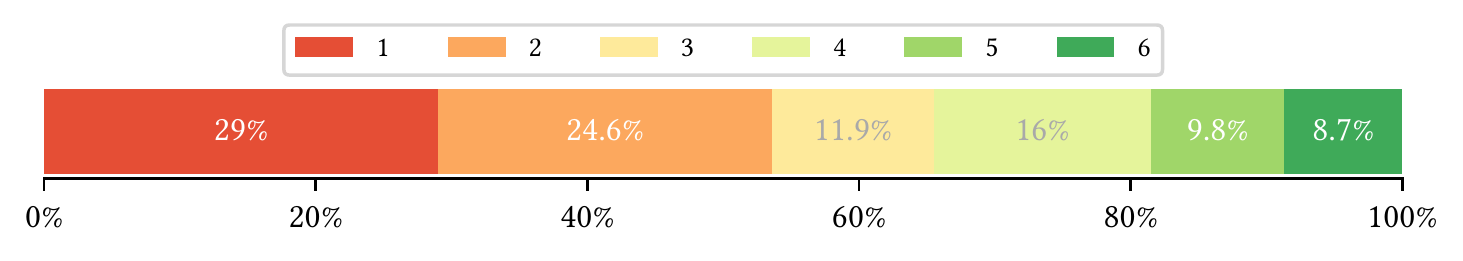}
    \caption{Distribution of the number of changes from single change (at the very left) to six changes (at the very right).}
    \label{fig:change-dist}
\end{figure*}

As previously said, the number of changed language tokens between the buggy and the fixed version of a dataset instance ranges between 1 and 6. The number of instances with a specific change count decreases as the number of changes increases. While in almost one third of cases there is a difference of a single token we see changes with six tokens in only 8.6\% of the data. The median number of changes is 2 for all languages except JavaScript and Python for which the median is 3. Figure~\ref{fig:change-dist} shows how the number of changed tokens distributes in the data.
We observe a similar distribution for line hunks, that is the number of consecutive blocks of lines with changes. Roughly 75\% of bugs are single-hunk bugs. 20\% of changes include two hunks, only 5\% more than two hunks (i.e., 3-6 hunks).

\vspace{0.8\baselineskip plus 0.2\baselineskip minus 0.2\baselineskip}
\begin{answerbox}{Answer to RQ1.2}
Almost 30\% of bugs in the dataset require only a single token change to fix. The higher the number of changes, the lower the proportion of data with that number of changes. Only 25\% of bugs require multi-hunk changes.
\end{answerbox}

\subsubsection{Bug Labels (RQ1.3)}

As described earlier, we use a rule-based method to automatically label all bugs. A bug can have multiple labels and labels are hierarchical. For instance, if, to fix a bug, it is necessary to change an if-condition, the label \texttt{control\_flow.branch.if.condition.change} is used. Similar labels share common prefixes. The removal of the \texttt{else} branch of an if-else will be labeled with \texttt{control\_flow.branch.else.remove} as it potentially changes control flow and branching. Adding a return statement, for example, similarly, alters control flow. It is, however, not related to branching: the used label is \texttt{control\_flow.return.add}.
Labels end with one or more actions, indicating the type of change. This action is either \texttt{.change}, \texttt{.add} or \texttt{.remove}.
A full list of labels is given in the appendix in the form of a diagram (Figure~\ref{fig:labels-diag}).
 
\paragraph{Most frequent labels}
Table~\ref{tab:labels} provides an overview of the most frequent labels grouped by language. There is relatively little variation across languages. In six out of the eight languages, bugs that require changing function or method call arguments are at the top (\texttt{call.arguments.change}). In the other two languages (C++ and Go), changes to if-conditions are most frequent. Overall, this type of bug is very common and is among the top-four most frequent change types for all languages. These results are consistent with previous studies. In particular, \citet{panUnderstandingBugFix2009}, after analyzing bug fix patterns in seven large Java open-source projects found fix patterns relating to method calls and if-conditions to be the most common (\enquote{method call with different actual parameter values} and \enquote{change in if conditional}). 

The label \texttt{io.output.change} is among the six most frequent labels for all but one language (Ruby). This label is assigned if changes are made that alter a programs output (e.g., changing a print format string or adding or removing a \texttt{print()}/\texttt{printf()}/\texttt{puts()} call).
Changes to literals (integers or strings) are also common, appearing in the top six list for five languages.
Finally, we observe that the most frequently modified operators are comparison operators (\texttt{expression.operator.compare.change}).
For an overview of all labels please see Figure~\ref{fig:labels-diag} in the appendix.

\answerspace{}
\begin{answerbox}{Answer to RQ1.3}
Changes to method and function call arguments and if-conditions are the dominating bug types in all languages. This is in line with previous work. Also frequent are changes to literals as well as output-altering changes (e.g., print calls or format string changes). The most commonly changed operators are comparison operators (e.g., \texttt{<},\texttt{>=}, \texttt{!=}).
\end{answerbox}

\begin{table}[htbp]
    \footnotesize
    \centering
    \caption{Most frequent bug labels by language. Frequency is the relative number of bugs associated with a specific label. As multiple labels are possible, frequencies do not to sum to 100\%.}
    \label{tab:labels}

    \begin{tabular}{cTr}
         \textbf{Language} & {\normalfont\bfseries Label} & \textbf{Freq.} \\
\toprule

\multirow{6}{*}{C} & call.arguments.change & 41.66\%\\
 & literal.string.change & 35.26\%\\
 & io.output.change & 32.98\%\\
 & control\_flow.branch.if.condition.change & 19.62\%\\
 & expression.operation.binary.change & 15.06\%\\
 & literal.number.change & 14.92\%\\
\hline
\multirow{6}{*}{C++} & control\_flow.branch.if.condition.change & 21.41\%\\
 & io.output.change & 14.96\%\\
 & literal.number.change & 14.96\%\\
 & expression.operation.binary.change & 14.56\%\\
 & expression.operator.compare.change & 12.92\%\\
 & control\_flow.loop.for.condition.change & 12.09\%\\
\hline
\multirow{6}{*}{JavaScript} & call.arguments.change & 37.90\%\\
 & io.output.change & 21.13\%\\
 & expression.operation.binary.change & 19.74\%\\
 & control\_flow.branch.if.condition.change & 17.35\%\\
 & literal.string.change & 15.62\%\\
 & identifier.change & 14.85\%\\
\hline
\multirow{6}{*}{Java} & call.arguments.change & 23.81\%\\
 & control\_flow.branch.if.condition.change & 21.02\%\\
 & expression.operation.binary.change & 19.51\%\\
 & io.output.change & 17.96\%\\
 & identifier.change & 16.64\%\\
 & expression.operator.compare.change & 15.24\%\\
\hline
\multirow{6}{*}{Ruby} & call.arguments.change & 25.60\%\\
 & control\_flow.branch.if.condition.change & 19.45\%\\
 & assignment.value.change & 16.54\%\\
 & call.add & 16.50\%\\
 & identifier.change & 15.85\%\\
 & literal.string.change & 11.52\%\\
\hline
\multirow{6}{*}{Python} & call.arguments.change & 37.51\%\\
 & control\_flow.branch.if.condition.change & 19.60\%\\
 & identifier.change & 16.31\%\\
 & io.output.change & 15.49\%\\
 & assignment.value.change & 15.34\%\\
 & expression.operation.binary.change & 13.59\%\\
\hline
\multirow{6}{*}{PHP} & call.arguments.change & 19.82\%\\
 & control\_flow.branch.if.condition.change & 19.72\%\\
 & io.output.change & 18.65\%\\
 & assignment.value.change & 17.43\%\\
 & literal.string.change & 15.55\%\\
 & expression.operation.binary.change & 11.79\%\\
\hline
\multirow{6}{*}{Go} & control\_flow.branch.if.condition.change & 24.47\%\\
 & call.arguments.change & 24.13\%\\
 & expression.operator.compare.change & 19.19\%\\
 & expression.operation.binary.change & 19.12\%\\
 & io.output.change & 16.87\%\\
 & control\_flow.loop.for.condition.change & 16.69\%\\

 \bottomrule
 
    \end{tabular}
\end{table}

\subsubsection{Errors \& Exceptions (RQ1.4)}

\paragraph{What is an error?}
\begin{figure*}[htbp]
	\begin{minipage}{0.45\textwidth}
		\begin{lstlisting}[language=Ruby,style=patch,numbers=left,xleftmargin=0pt,xrightmargin=0pt,framesep=0pt]
def is_triangle(x, y, z):
    x, y, z = sorted((*\btHL[fill=green!30]{[}*)x, y, z(*\btHL[fill=green!30]{]}*))

return x**2 + y**2 == z**2

num = int(input())
for i in range(num):
    x, y, z = map(int, input().split())
    triangle = is_triangle(x, y, z)
    if triangle:
       print('YES')
    else:
        print('NO')
		\end{lstlisting}
	\end{minipage}
	\hfill
	\begin{minipage}{0.45\textwidth}
		\footnotesize
	\begin{lstlisting}[basicstyle=\ttfamily,morekeywords={Traceback,File,line,TypeError}]
Traceback (most recent call last):
  File "file.py", line 8, in <module>
    triangle = is_triangle(x, y, z)
  File "file.py", line 2, in is_triangle
    x, y, z = sorted(x, y, z)
  TypeError: sorted expected 1 argument, got 3
\end{lstlisting}

	\end{minipage}
	\caption{The \texttt{sorted} function takes a single argument of type \texttt{list} (bug \#885329). Because multiple arguments are passed, a \texttt{TypeError} is raised. The stack trace (on the right) immediately reveals the bug location.}
	\label{fig:example-bug-error}
\end{figure*}

As stated earlier, we run all dataset programs on all test inputs.
A program might either \begin{inparaenum}[a)]
\item output the correct answer,
\item output a wrong answer,
\item timeout or
\item produce a runtime error.
\end{inparaenum}
It is not always clear-cut whether an error occurred during program execution.
We resort to a simple heuristic and declare an execution to have produced an error if either
\begin{inparaenum}[a)]
\item the execution terminates because of an operating system-level signal (e.g., \texttt{SIGSEV} or \texttt{SIGABORT}) or
\item the program outputs to standard error (i.e., \texttt{stderr}) \emph{and} indicates failure by returning a non-zero exit status.
\end{inparaenum}
Programs may produce an error and still output the correct answer to standard output. In this case the correct output is ignored.
For languages that consistently handle errors with exceptions we are able to extract the exception name and to calculate exact frequencies of occurrence. Such languages include Python, Ruby and Java.
For C++ and JavaScript, errors might not always contain an exception type. We report all exceptions found in error messages.
The PHP interpreter outputs warnings and errors to standard output instead of standard error. This makes determining errors difficult and unreliable. Our analysis of PHP errors is thus limited to matching program output against common patterns of error messages.
Determining errors is more reliable for C/C++ and Go. However, because these languages do not have an exception system (or in the case of C++ often fail without throwing an exception) we likewise resort to matching error messages against a set of common message patterns.
Figure~\ref{fig:example-bug-error} shows a Python bug in our dataset that raises a \texttt{TypeError}. Table~\ref{tab:exceptions} lists the most common exceptions that were thrown during execution, along with relative frequencies.

\paragraph{Dynamic languages}
For Python, Ruby and JavaScript, the most frequent exception types are related to the dynamic nature of these languages. Such errors include \begin{inparaenum}[a)]
    \item referencing undefined variables, methods, functions or attributes (\texttt{NameError}, \texttt{ReferenceError}, \texttt{NoMethodError}, \texttt{AttributeError}) and
    \item type errors (\texttt{TypeError}).
\end{inparaenum}
In Python, similar to Java, index and key errors caused by out-of-bounds access of arrays or maps/dictionary lookups with invalid keys are relatively frequent (roughly 17\%), while they occur less often in Ruby and JavaScript (less than 1\%). The probable reason for this difference is likely that Python raises an exception for out-of-bounds accesses and lookups while Ruby and JavaScript return a null value (\texttt{nil} in Ruby, \texttt{undefined} in JavaScript). For PHP we identify a similar pattern: the most frequent error message is \enquote{\texttt{Use of undefined constant}} caused, for instance, by referencing undefined variables. Argument errors are another frequent group of errors in PHP programs (caused by, e.g., passing the wrong number or the wrong type of arguments to function calls). Unfortunately, due to above mentioned limitations, we cannot reliably quantify error frequencies for PHP.

\paragraph{Static languages}
C/C++, Go and Java are all static languages. Name errors are caught during compilation and thus cannot occur at runtime. For Java, C++ and Go we find errors related to out-of-bounds access to be the dominating type. In Java, the top three exception types are all related to invalid bounds and account for over 50\% of all thrown exceptions. Similarly, for C++ \texttt{std::out\_of\_range} is the most frequently thrown exception. Since array access in C/C++ is very often unchecked (in which case the program crashes without throwing an exception) the number of bounds-related bugs might actually be higher. Finally, error messages for Go paint a similar picture. Here, 71\% of error messages matched the pattern \texttt{index out of range} and 6\% \texttt{slice bounds out of range}.

With only 2.59\%, the frequency of \texttt{NullPointerException}s occurring during execution of Java programs is surprisingly low; previous work found that in Android applications up to 27.7\% of crashes are related to this type of error~\citep{coelho2017exception}.

\paragraph{Memory errors} C and C++ are unsafe languages with manual memory management. As expected, we see a large number of memory related issues during execution of programs written in these languages. For C++, we find keywords such as \texttt{malloc}, \texttt{free()}, \texttt{double free}, \texttt{invalid pointer} or \texttt{stack smashing} in over 30\% of cases. This is even higher for C, where over 90\% of error messages contain one of the above keywords.
However, since C does not have an exception system and memory issues are among the few errors that are reported to standard error, these numbers are to be taken with a grain of salt. Since memory access is usually not checked in C (and not always in C++) memory errors can have various causes, among others, also invalid bounds.

\answerspace{}
\begin{answerbox}{Answer to RQ1.4}
The types of errors and exceptions strongly depend on whether a language uses static or dynamic typing. For dynamic languages (Python, JavaScript, Ruby, PHP) variable or method name resolution errors are most frequent. For static languages (C/C++, Java, Go) the majority of errors are related to invalid bounds. Memory errors are common for C++ and especially C.
As far as dynamic languages are concerned, we notice that the line between syntactic and semantic program repair is rather blurry as many static-language errors (i.e., compiler errors) handled by syntactic repair here occur at runtime (e.g., use of an undefined variable, method or function).
\end{answerbox}

\begin{table}[htbp]
    \footnotesize
    \caption{Most frequent exceptions thrown during execution. Frequencies are counted per program/bug.}    
    \centering
    \label{tab:exceptions}
    \begin{threeparttable}
    \begin{tabular}{cTr}
         \textbf{Language} & {\normalfont\bfseries Exception/Error} & \textbf{Freq.} \\
\toprule

\multirow{10}{*}{Python} & NameError & 33.28\%\\
& TypeError & 25.77\%\\
& IndexError & 16.64\%\\
& ValueError & 9.68\%\\
& AttributeError & 5.57\%\\
& ZeroDivisionError & 2.69\%\\
& EOFError & 2.22\%\\
& SyntaxError & 1.21\%\\
& KeyError & 0.79\%\\
& ModuleNotFoundError & 0.79\%\\
\hline
\multirow{10}{*}{Java} & ArrayIndexOutOfBoundsException\tnote{1} & 36.99\%\\
& InputMismatchException\tnote{2} & 15.46\%\\
& StringIndexOutOfBoundsException\tnote{1} & 9.59\%\\
& NoSuchElementException\tnote{2} & 8.81\%\\
& ArithmeticException\tnote{1} & 5.23\%\\
& NumberFormatException\tnote{1} & 5.09\%\\
& ClassNotFoundException\tnote{1} & 3.03\%\\
& IndexOutOfBoundsException\tnote{1} & 2.89\%\\
& NullPointerException\tnote{1} & 2.59\%\\
& NegativeArraySizeException\tnote{1} & 2.05\%\\
\hline
\multirow{10}{*}{Ruby} & NoMethodError & 48.11\%\\
& NameError & 27.61\%\\
& TypeError & 11.44\%\\
& ArgumentError & 6.28\%\\
& ZeroDivisionError & 2.54\%\\
& LoadError & 1.07\%\\
& IndexError & 0.68\%\\
& SystemStackError & 0.40\%\\
& FloatDomainError & 0.32\%\\
& NoMemoryError & 0.24\%\\
\hline
\multirow{4}{*}{JavaScript} & ReferenceError & 51.94\%\\
& TypeError & 39.72\%\\
& RangeError & 1.39\%\\
& SyntaxError & 0.83\%\\
\hline
\multirow{8}{*}{C++} & std::out\_of\_range & 32.62\%\\
& std::length\_error & 11.39\%\\
& std::bad\_alloc & 10.35\%\\
& std::invalid\_argument & 1.07\%\\
& std::logic\_error & 0.42\%\\
& std::bad\_array\_new\_length & 0.32\%\\
& std::runtime\_error & 0.03\%\\
& std::bad\_function\_call & 0.03\%\\
\hline
    \end{tabular}
\begin{tablenotes}
\item[1] in \texttt{java.lang}; \item[2] in \texttt{java.util}
\end{tablenotes}
\end{threeparttable}
\end{table}

\subsection{Baselines (RQ2)}

As previously stated, we evaluate the dataset on a NPR and a G\&V baseline. The neural baseline is based on CodeT5~\citep{codet5}. For the G\&V baseline we rely on Cardumen~\citep{cardumen}. The goal of this evaluation is to provide future users of this dataset with a point of reference with respect to the performance they can expect from their tools.

\subsubsection{Evaluation}\label{sec:baseline-eval}

For the NPR baseline, we used HuggingFace's \texttt{transformers} library~\citep{huggingface} to fine-tune the publicly available \texttt{codet5-small} model checkpoint on the training set. We frame the repair process as a seq2seq translation problem (from buggy to fixed), that is, the model learns to jointly \emph{localize and repair} defects. Consequently, we do not use bug location markers (like e.g., SequenceR~\citep{chenSequenceRSequencetoSequenceLearning2019}) nor assume perfect localization.
We train for five epochs using a learning rate of 1e-4, two gradient accumulation steps and a batch size of 16 on a single NVIDIA RTX 3090 with 24GB of memory.
For inference (i.e., test set evaluation), if not specified otherwise, we produce five outputs per instance (i.e., five candidates) employing beam search with six beams. Previous work suggest that for the task at hand beam search should be preferred over greedy decoding~\citep{haqueFixEvalExecutionbasedEvaluation2022}. To estimate the effect of the number of candidates on performance, for a single experiment we generate up to ten candidates per bug (Figure~\ref{fig:candidate-perf}). 
During training and inference we prefix all submission with a special language identification token (e.g., \texttt{<python>}). This idea is taken from PLBART~\citep{plbart}, a successful multi-language code model.

Cardumen can only deal with bugs in Java code and requires JUnit test cases. We automatically translate all test cases into JUnit4 tests and only test on Java bugs in the test set. In order to be able to run Cardumen on thousands of programs we had to reduce the maximum running time to 10 minutes and the number of generations to 100 (\texttt{-maxtime 10 -maxgen 100}). This is an unusually low time budget for a G\&V tool where time budgets are often in the range of hours~\citep{cardumen}. The long running time of G\&V systems has been raised as a issue in previous work~\citep{nodaExperienceReportHow2020a}. It not only limits the practicality of G\&V approaches, but also makes it very difficult to evaluate these tools on a large number of bugs.

For fault localization we rely on ASTOR's\footnote{ASTOR~\citep{martinez2016astor} is the framework surrounding Cardumen}  new Flacoco~\citep{flacoco} fault localization engine instead of GZoltar~\citep{campos2012gzoltar}, as the latter is incompatible with recent Java versions.

\subsubsection{Baseline Performance (RQ2.1)}

\paragraph{G\&V baseline} Cardumen successfully (i.e., plausibly) fixed 157 of the roughly 3,300 Java bugs. This result is rather disappointing. However, as was stated previously, in order to handle such a large number of bugs, we had to cut the time budget to an unusually low value. Also note that G\&V systems operate under the redundancy assumption, that is, the assumption that the correct ingredients for a fix can be found in the program under repair. Since here we are dealing with very small programs, there is little or no redundancy. This problem has been discussed in previous work~\citep{comprehensivequixbugs}.

\paragraph{NPR baseline}
The NPR model based on CodeT5 performed surprisingly well and plausibly repaired  59.8\% of submissions (across all languages with five candidates per bug).



\paragraph{Performance \& number of changes}
We calculate Pearson's r between the number of changes (changed tokens) and baseline performance (plausibility rate). As expected, we find a significant negative correlation for both baselines, that is, performance decreases with an increasing number of changes. With $r = -0.996$ this correlation is stronger for the neural baseline that for the G\&V baseline where $r = -0.6823$.
Performance of the neural baseline ranges from 68.7\% for single change fixes to 42.8\% for fixes that require six token changes (with five candidates per bug); for the G\&V baseline these numbers are 2\% and 0.09\%, respectively (on Java bugs only).


\paragraph{Performance across languages}
Cardumen, our G\&V baseline is a Java-only tool. We thus focus on our NPR baseline. First, with $r = 0.33$ we see a moderate correlation between the number of training samples and performance for a particular language. Performance (i.e., plausibility) with five candidates per bug is best for Python (69\%), followed by C (64\%) and Java (62\%); the hardest languages were Ruby (47.6\%) and Go (47.8\%). Please refer to Table~\ref{tab:npr-results} for more details.

\answerspace{}
\begin{answerbox}{Answer to RQ2.1}
With barely 5\% of plausibly fixed bugs, Cardumen, the G\&V baseline, fared rather poorly. In contrast, the NPR model based on CodeT5 performed surprisingly well and plausibly fixed close to 60\% of bugs.
As expected, performance is lower for more complex bugs (higher token change count). Performance is best (> 60\%) for C, Python and Java while Ruby and Go bring up the rear (< 50\%).
\end{answerbox}

\subsubsection{Performance across Labels -- Strengths \& Weaknesses (RQ2.2)}

Bug labels allow us to perform a per-label analysis. We consider two types of analyses. First, we calculate per-label plausiblity rate. This shows the rate at which specific labels are fixed. Second, we provide a weak and strong point analysis. Here, we calculate the plausiblity of a specific label \textit{relative} to its frequency in the dataset. We call this the $r$ value of a label. The corresponding section below describes in more detail how $r$ values were obtained.

\paragraph{Performance across Labels}

We start by looking at the overall performance of CodeT5 over each label with a support of 30 or more in the test set (specifically, the percentage of plausibly fixed bugs that have a given label), and aggregate it into categories. On fine-grained labels, the performance ranges from 90\% or more (adding unary references, adding php tags), to less than 20-25\% (adding continue keyword or changing some operators), showing ample variations between different types of bugs.  If we aggregate into our labels into top-level categories (Tabel~\ref{tabl:per-label-agg}), we see that categories related to variables (identifiers, 52\%; assignments, 48\%; variable access, 40\%) seem to be underperforming (with the exception of variable declarations, 66\%). Function return values and preprocessor categories also underperform (34\%). The \texttt{control\_flow} category (58\%) is slightly worse than the average over all bugs (60\%), while function calls (63\%), literals (66\%), input/output (76\%), and type conversion (81\%), on the other hand, are overperforming. Of note, the bugs that are as yet unlabelled have a performance slightly lower than the average (53\%), indicating that the \enquote{long tail} of unlabelled bug is more challenging overall.

Note that the performance is not very correlated to the amount of data (assuming the distribution on the train set is equivalent). It is only 0.09, indicating that others factors are at play, such as the inherent difficulty of individual labels (e.g. \texttt{php\_tag} is the highest performing label, but is somewhat rare). 

For Cardumen, we see that many labels have no plausibly fixed bug at all. These labels include adding function or method calls, adding missing newlines to strings or adding type conversions. The most successful labels were related to operators. Removing operators had the highest plausibilty rate at almost 12\%, changing arithmetic operators 9\%. 

Next, we formalize the analysis to detect weak and strong spots, taking into account the size of the data.

\begin{table}[htbp]
\caption{Per-label performance (plausiblity rate) of our NPR baseline aggregated into groups with macro and micro average, respectively.}
\centering
\begin{tabular}{@{}rcc@{}}
                     & Macro & Micro \\ \midrule
Type Conversion      & 78.6\%      & 81.3\%      \\
IO                   & 75.5\%      & 75.5\%      \\
Literal              & 51.8\%      & 66.1\%      \\
Variable Declaration & 57.0\%      & 65.5\%      \\
Call                 & 64.2\%      & 63.3\%      \\
All                  & 59.8\%      & 59.8\%      \\
Misc                 & 70.0\%      & 59.4\%      \\
Expression           & 55.1\%      & 59.3\%      \\
Control Flow         & 50.8\%      & 57.8\%      \\
No label             & 52.5\%      & 52.5\%      \\
Identifier           & 49.8\%      & 51.9\%      \\
Assignment           & 48.0\%      & 47.9\%      \\
Variable Access      & 39.6\%      & 39.6\%      \\
Function             & 34.1\%      & 34.1\%      \\
Preprocessor         & 33.7\%      & 33.7\%      \\ \bottomrule
\label{tabl:per-label-agg}

\end{tabular}
\end{table}

\paragraph{Strengths \& Weaknesses}

We use the following approach to detect such weak and strong spots. Let $f(L, l)$ be the relative frequency of label $l$ relative to labels in the multiset $L$. Then, for each label $l$ of a plausibly fixed bug (from the test set) we calculate the quotient $$r_l = \frac{f(L_{plausible}, l)}{f(L_{test}, l)}$$ where $L_{test}$ is the multiset of all labels in the test set and $L_{plausible}$ is the multiset of all labels in the set of plausibly fixed bugs. We sort all labels in $L_{plausible}$ according to $r_l$. We consider the top-$k$ (we use $k=5$ in our analysis) labels \enquote{strong spots}, the bottom $k$ labels as \enquote{weak spots}. Generally, $r \gg 1$ indicates a strong spot, while $r \ll 1$ a weak spot. To get reasonably reliable numbers, we calculate and analyze $r_l$ only for labels with an absolute frequency in the test set of 30 or more.

\paragraph{Strong points}
For our NPR baseline the label with the highest $r$-score
is \labeltt{expression.operation.unary.reference.add}. This label is assigned if fixing the bug involved adding a reference operator in C, C++, Go, e.g., \texttt{scanf("\%d", {\btHL[fill=green!30]\&}mnt[i]);}.
Other strong points are:
\begin{itemize}
    \item changing the case of string labels (e.g. \texttt{System.out.println(m + {\btHL[fill=red!30]"X"}{\btHL[fill=green!30]"x"} + x + "=" + x * m)},
    \item adding missing PHP tags (\texttt{\btHL[fill=green!30]<?php}),
    \item replacing an assignment operator with a comparison operator (e.g., \texttt{if (a[1] {\btHL[fill=red!30]=}{\btHL[fill=green!30]==} 3)}),
    \item adding a type conversion (e.g., \texttt{cout <{}< {\btHL[fill=green!30](int)}money <{}< endl;)},
    \item changing variable declaration types (e.g., \texttt{{\btHL[fill=red!30]int}{\btHL[fill=green!30]long long} a, b;}) and finally
    \item changing an if-branch to an else-if-branch (\texttt{{\btHL[fill=green!30]el}if y == base:}).
\end{itemize}

For Cardumen, bug fixes involving an operator inversion where fixed most successfully,
e.g., changing \texttt{{\btHL[fill=red!30]<}} to \texttt{{\btHL[fill=green!30]>}}, \texttt{{\btHL[fill=red!30]!=}} to \texttt{{\btHL[fill=green!30]==}} etc., followed by bugs that require the removal, replacement or change of simple binary operations (e.g., \texttt{long remainder = x \% 11 {\btHL[fill=red!30]* 2};}, \texttt{System.out.println({\btHL[fill=red!30]n * a}{\btHL[fill=green!30]1});} and 
\texttt{System.out.println(10 * b {\btHL[fill=red!30]*}{\btHL[fill=green!30]+} a + c);}).

\paragraph{Weak Points}
The largest weak spot of the NPR baseline is adding if-branches, 
(e.g., \texttt{\btHL[fill=green!30]if j < m:}), as well as adding missing \texttt{\btHL[fill=green!30]continue}s in loops. 
While the label for changing string literals is among the 15 labels with the highest $r$, the model seems to struggle with integer literals, 
likely because such changes are often quite specific without general rules (e.g., \texttt{\#define N 200000{\btHL[fill=red!30]0}{\btHL[fill=green!30]2}}). Changes of integer literals often occur inside array dimension declarations or assignments (e.g., \texttt{int P[{\btHL[fill=red!30]10}{\btHL[fill=green!30]11}];}). Labels for these two changes, that is, adding, changing or removing assignments or changing array dimensions had similarly low $r$ values.

Due to the relatively bad performance of Cardumen, many labels have no fix at all. We see, for instance, that Cardumen failed to fix any bug that requires adding a method or function call. Among the plausibly fixed bugs there were also none that involved the addition of a type conversion. It also showed bad performance fixing buggy for-loop initializers 
or conditions.

\answerspace{}
\begin{answerbox}{Answer to RQ2.2}
The CodeT5 model is good at fixing specific operator bugs (missing reference operator or use of \texttt{=} instead of \texttt{==}) as well as numeric type conversions and declarations of the wrong (numeric) type. The model seems to struggle with certain control flow bugs requiring the insertion of a \texttt{continue} keyword or a new if-branch. 
Most of the plausibly fixed bugs by Cardumen involve some form of simple expression modification, such as inverting comparison operators or swapping arithmetic operators. Outside this class of simple bugs, Cardumen has trouble with many types of bugs, for instance it fails to fix any bug requiring the addition of a function call or a type conversion.
\end{answerbox}

\subsubsection{Knowledge Transfer between Languages (RQ2.3)}

\begin{table}[htbp]
    \caption{Results (plausiblity rate in percent) of the NPR (CodeT5) baseline with different training regimens: training was done on all languages (all), all languages except the test language (no) and only the target language (only). All numbers have been rounded to a single decimal place and have been obtained using five candidates per bug.}
    \centering
    \label{tab:npr-results}
    \begin{tabular}{rrrrr}
        \footnotesize

         \textbf{Language} & \textbf{All} & \textbf{No} &\textbf{Only} & \textbf{Transfer} \\
         \toprule
         C & \textbf{64.4} & 60.6 & 50.0 & +14.4 \\
         C++ & \textbf{54.9} & 26.3 & 53.7 & +1.2 \\
         Java & \textbf{62.9} & 50.8 & 40.9 & +22.0 \\
         Python & \textbf{69.0} & 26.2 & 66.8 & +2.2 \\
         Ruby & \textbf{47.6} & 13.7 & 34.0 & +13.6 \\
         Go & \textbf{47.8} & 1.1 & 16.0 & +31.8 \\
         PHP & \textbf{55.7} & 24.2 & 21.2 & +34.5 \\         
         JavaScript & \textbf{44.5} & 34.4 & 16.2 & +28.3 \\

 \bottomrule
 
    \end{tabular}
\end{table}

Generally speaking, knowledge transfer occurs when a model is able to apply to a task knowledge it learned from another task. This strategy is particularly useful in cases where data is scarce for a particular task but not for a similar other task. For instance, in order to improve machine translation performance for languages with low data-availability \citet{zoph-etal-2016-transfer} first trained the model on a related  \enquote{high-resource language} (i.e., a language with high data availability). Recent work suggests that a similar approach can be viable for low-resource programming languages in a variety of software engineering tasks (e.g., code summarization or function naming)~ \citep{multilingual-training}.
Here, we are interested in the question of whether bugs and their fixes can be transferred from one language to another. In particular, we want to find out if models for fixing bugs in languages with little data (e.g., PHP or Go) can benefit from training on other languages with copious data (e.g., C++ or Python).

To estimate this knowledge transfer effect between languages we train 17 NPR models with a different training regimen. One model is trained on the entire training set (i.e., all languages). Then, for each of the eight languages $l$, we train two more models, one with \emph{no} training instances of language $l$ and one with \emph{only} instances of languages $l$. Results are shown in Table~\ref{tab:npr-results}.
The \textit{language} column specifies the test language, the \textit{all} column shows the result when trained with the entire training set (all languages), the \textit{no} column when trained on all training instances \emph{not} in the test language and the \textit{only} column when trained only on training instances of the test language. E.g., when the test language is JavaScript, the \textit{no} column shows performance when trained on all training data samples \emph{not} in JavaScript while the \textit{only} column shows performance when trained on JavaScript training samples only.

Across all languages, the plausibility rate is higher on the model trained on all languages, indicating that there is some form on knowledge transfer in all cases.
We see that languages with small training sets benefit most from cross-language training (Go, PHP, JavaScript). Conversely, for languages with large training sets (C++, Python) this gain is much smaller (1-2\%).

For Java, C, PHP and JavaScript training on all other languages yields higher results than training only on the language itself. For PHP and JavaScript this is likely attributable to their very small training sets. For Java and C we conjecture that this might be due to the languages' similarity to C++ (i.e., similar syntax and types). Using the example of Java, given this similarity, the model might learn more bugfix patterns from over 250,000 C/C++ bugs (plus other non-Java languages) than from 13,000 Java bugs, despite the language difference.

\answerspace{}
\begin{answerbox}{Answer to RQ2.3}
We clearly see some knowledge transfer between different programming languages. This transfer effect is larger for languages with small training sets. This result provides preliminary evidence that languages with low data-availability can benefit from training data in different languages.
\end{answerbox}

\subsubsection{Number of Candidates in NPR (RQ 2.4)}

\begin{figure*}
    \centering
	\includegraphics[width=\linewidth]{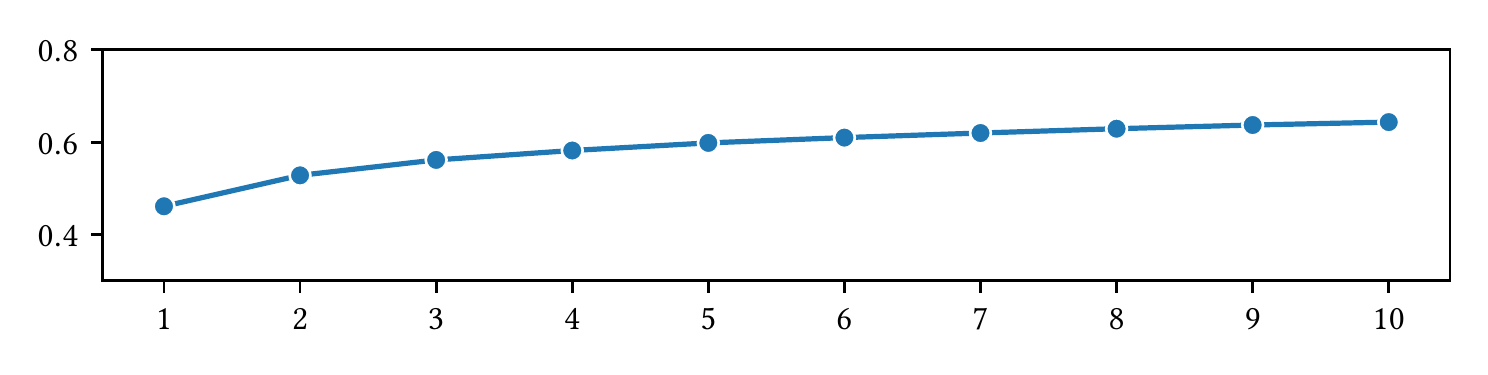}
    \caption{Plausibility rate (y) of the CodeT5-based model for different numbers of candidates per bug (x) when trained and evaluated on all languages.}
    \label{fig:candidate-perf}
\end{figure*}

In our experiments with the NPR baseline we generated five candidates per bug. This number tries to balance the computational cost of code generation with a decent number of candidates. However, the question inevitably arises how performance would change if we generated less or more candidates. To answer this question, for the model trained on all languages, we generate and evaluate up to 10 candidates. Figure~\ref{fig:candidate-perf} shows performance (plausiblity rate) for 1 to 10 candidates per bug. With a single candidate, performance drops to 46.13\%, while it raises to 64.33\% with 10 candidates. However, doubling the computational cost from 5 to 10 candidates yields a \emph{relative} performance improvement of only 7.5\%.
Our online supplement contains a graphical visualization of candidate evaluation results (see Table \ref{tab:candidate-vis} in the appendix).

\answerspace{}
\begin{answerbox}{Answer to RQ2.4}
When generating a single candidate per bug, the overall plausibility rate is only 46\%. This number increases to over 64\% with 10 candidates per bug. However, relative improvement diminishes as the candidate count is increased.
\end{answerbox}

\section{Discussion}\label{sec:discussion}

Based on the results of the baselines, we can conclude that there is ample room for improvement, with different challenges to address for each kind of approaches.

\paragraph{Challenges for G\&V approaches.} The performance of our G\&V baseline, Cardumen, is rather disappointing. However there are at least two factors to take into account: the time budget was rather constrained, and the search space for ingredients was very limited due to low redundancy in submissions. Relaxing both constraints would probably improve performance. The time constraint emphasizes an existing issue with G\&V approaches: they can be very slow, which hampers their evaluation at scale. Taken together, these two factors point toward the need for more efficient search approaches in APR, if they are to be competitive with NPR approaches. The second challenge for G\&V approaches is their language-specificity: we could only run Cardumen on the Java portion of our benchmark. This did alleviate the scaling issue as it reduced the size of the test set by an order of magnitude. However the lack of language independence is still a major limitation, compared to the straightforward support this has in NPR approaches.

\paragraph{Challenges for NPR approaches.} We were pleasantly surprised by the performance of the NPR baseline. This showed both that the benchmark is at an accessible level of difficulty, but also that there is significant room for improvement: more than 40\% (35\% with 10 candidates) of the bugs can not be repaired by our straightforward baseline. More than that, we can expect that the 40\% of remaining bugs will be more challenging. The performance of the NPR approach is very correlated with the size of the bug fix (much more so than the G\&V approach), indicating that the NPR approach works well for the ``simpler'' bugs, but not so well for the more complex bugs. This is also reflected in the labels were the model was the most and least successful: the strongest points were simpler fixes to, e.g., operator usage, while the weakest points concerned issues relating to the program's control flow.

\paragraph{Leveraging transfer learning.} We found very encouraging results from transfer learning. These results explain the lower than expected negative correlation between training set size and performance: a small training set is not as much of an issue when it is possible to learn from similar languages (or similar problems in other languages), and it opens up possibilities to apply NPR techniques to languages that are less popular. Thus, further research should be conducted in defining models that exploit data from other languages more efficiently than our baseline does. On the other hand, perhaps there is a limit to the effectiveness of transfer learning: it is possible that for bugs that are more complex, or that use more exotic language features (e.g., keyword arguments in Ruby), transfer learning will be less effective.

\paragraph{Leveraging execution data.} So far, we have provided performance numbers for baselines that leverage static data only. We have not used at all the execution data that we have collected as a learning signal. We think that exploiting this signal would significantly improve model performance. Simply making available to the model the type of runtime error to fix (e.g., a NullPointerException) would be a very precious hint as it should scope down the problem. Similarly, knowing that the issue is \emph{not} a runtime error but a different issue (e.g., a wrong calculation) would also help to scope down the problem. Further down the line, knowing more information about failing test cases could be very valuable to NPR models.

\paragraph{Towards hybrid approaches?} Another avenue that would leverage execution data would be to provide feedback to the model about the repair attempts. Approaches such as RewardRepair~\cite{yeNeuralProgramRepair2022} are first steps in this direction; however RewardRepair is limited by the amount of training data with actual execution information. One could imagine multiple variants of these approaches. These could constitute hybrids between classical G\&V approaches and NPR approaches: an NPR approach could generates patches, which are executed to ensure they are plausible; if not,Wird nicht funktionieren.
 the NPR approach could be called multiple times until the test passes. ARJANMT~\citep{arjanmt}, a
G\&V-NPR hybrid, for instance, uses an NPR model as source for further ingredients. Allowing such systems to further exploit execution information could make them more efficient.
NPR systems could also be extended in other ways, for example by feeding them data from previous executions, perhaps using mechanisms similar to \enquote{scratch pads} \cite{ney2021showYourWork} or by allowing them to use tools, like the Toolformer~\cite{schick2023toolformer}.

\paragraph{Evolving the benchmark.}
Evolving RunBugRun is possible in the future. Should performance on the benchmark start to saturate (e.g. exceeding 95\%), progress will become more difficult to measure. If this happens, we see several ways to make the task more challenging:
\begin{itemize}
    \item RunBugRun features bug fixes ranging from 1 to 6 tokens in size. We expect more rapid progress on smaller changes. An updated version could feature larger bug sizes instead, or reduce the weight of the smallest bug fixes.
    \item Similarly, we expect that some bug categories will be solved faster than others. An updated version of RunBugRun could filter out such categories, or increase the weights of the most challenging categories.
    \item A third possibility is to augment the number of programming languages, adding less popular languages for which there will be less data to learn from, to further incentivize models that leverage transfer learning. Another option is to simply limit the number of bugs in the training to incentivize more sample-efficient models.
    \item Finally, to place more emphasis on the ability of models to generalize to new problems, the way the benchmark is split could be changed, so that problems included in the test set are not included in the training set, or vice versa.
\end{itemize}

\section{Limitations, Biases and Threats to Validity}\label{sec:limitations}
\paragraph{Limited scope}
The dataset consists of short implementations of solutions to algorithmic problems.  There is a significant difference between the program code in the dataset and the code of larger software projects which might contain complex class hierarchies, UI-related code, large numbers of methods, functions and code files, dependencies on software frameworks and libraries, build scripts and so on. While there is some discrepancy between these two types of code, according to our findings the distribution of bug types in our dataset is similar to the distribution of bugs in open-source projects, as studied in previous work~\citep{panUnderstandingBugFix2009}. Moreover, as we have pointed out earlier, simple programs allow for very efficient execution and facilitate a large range of experiments. Also, many bugs in large projects are too complex for current APR systems (especially bugs that require multiple changes or that span multiple methods and files). Thus, we believe that the presented dataset is an interesting testbed for important next steps on the path to being able to tackle bugs in complex software systems.

\paragraph{Lack of diversity}
The test set contains programs solving 2,693 different problems. Because solution programs to the same problem inevitably bear a certain similarity the variety of code and consequently the variety of bugs is, to a certain degree, limited. We mitigate this issue by providing bug labels such that the different types of bugs and per-type performance can be better understood. 

\paragraph{Tokenization}
We implemented tokenizers for all eight languages. We tried to correctly tokenize a large portion of the corresponding language specification. However, perfectly tokenizing all eight languages according to its specification would require a huge engineering effort way beyond the scope of this work.

\paragraph{Labeling} To label bugs we manually created a set of rules that match certain patterns in the parse trees. Such rules are often language specific due to differences in the parse tree format or language features. We tried our best to cover a large number of changes. However, we cannot fully exclude that an import class of changes is not adequately captured. Unfortunately, existing solutions such as GumTree~\citep{falleri2014fine} do currently not support all languages in our dataset.
As existing tools are steadily improving and implementing support for more and more languages, we hope to be able switch to such a more robust and sophisticated tool in the future.

\paragraph{Plausibility vs correctness}
Manual patch correctness assessment is tedious and inefficient. With PATCH/TEST-SIM~\citep{xiong2018identifying}, DiffTGen~\citep{xin2017identifying} and RGT~\citep{ye2021automated} there are efforts to automate this task. Nonetheless, manual assessment is still common practice. With the size of the test set numbering in the thousands, however, manual assessment is no longer feasible with reasonable effort.

Many tests in our dataset were taken from the dataset used to train and evaluate AlphaCode~\citep{alphacode}, a deep learning model that is able to synthesize entire solution programs given a problem description. The authors estimate that 4\% of synthesized programs that pass the test suite are actually incorrect. We expected the portion of plausibly but incorrectly fixed programs to lie in a similar range. However, to get a more accurate estimate, we manually assessed a sample of 200 randomly selected plausible fixes generated by CodeT5 and found 7 (3.5\%) to be incorrect. 46.5\% of the fixes in the sample were identical to the ground-truth patch. For 19 patches, deciding correctness turned out to be very difficult. In that case, we ran the patched program with random input and compared the output to the output of the correct ground-truth program (on same input). If disagreeing input was found in a reasonable time span (1-2 minutes), we considered the patch invalid, otherwise valid. Our assessments are part of our online supplementary material\footnote{\url{https://giganticode.github.io/rbugr/assessment/}}. The rate of plausible but incorrect fixes reported in previous work is often higher. E.g., for CoCoNuT~\citet{lutellierCoCoNuTCombiningContextaware2020} authors report 35\% for QuixBugs~\citet{linQuixBugsMultilingualProgram2017} and 41\% for Codeflaws~\citep{tanCodeflawsProgrammingCompetition2017}. As previous work has shown~\citep{ye2021automated}, manual correctness assessment is prone to mistakes.
Should future research show specific problems to have a weak test suite, additional test cases can be generated by exploiting the fact that all programs can be executed.
Finally, the presented dataset may also serve as a starting point for further research on automatic patch correctness assessment.

\section{Concluding Remarks}\label{sec:conclusion}

In this work we presented RunBugRun, a large, multi-lingual dataset of executable bugs. Our baseline evaluation shows that there is still room for improvements, especially for G\&V approaches. Also, our results highlight that performance varies strongly across different programming languages as well as the number of changes. While our CodeT5 baseline could achieve a plausibility rate of 69\% for Python, performance for JavaScript, Go and Ruby was considerably lower (44.5-47.8\%). Generally, we observe that performance is lower for languages with lower data-availability. Fortunately, our experiments suggest that low-resource languages can benefit from knowledge transfer effects: mixing-in training data of popular languages available in abundance boosted performance  in all combinations we tried.
Performance also degrades with the number of changes that are required to fix the bug. Single change fixes generated by our NPR baseline were plausible with 68.7\%. However, this value drops for bugs that require multiple changes: a plausible fix was generated in 49\% of the cases for four or more changes and below 43\% for six changes. This indicates that there is still a long way to go before APR system can tackle complex bugs with dozens of changes and multiple hunks. Still, as APR tools become more advanced, in order to obtain a set of more challenging bugs the maximum change limit could be raised in the future (we only included bugs that can be fixed with less than seven token changes). 

This paper only marks the starting point of the RunBugRun dataset. We plan to continually expand it in several ways.
For instance, our dataset currently contains programs written in eight languages (C, C++, Python, Java, Ruby, JavaScript, Go, PHP). In the future we plan to add further languages.
We are particularly excited about future work exploiting runtime information. As all programs are executable, runtime traces can be easily obtain and used, for example, as sole or additional input to machine learning models.
We plan to build a parallel dataset of such execution traces. Figure~\ref{fig:execution-traces} shows an example trace obtained from a dataset instance in Python. In particular for fault localization, this form of information bears great potential. For instance, \citet{wang2017dynamic} have used neural embeddings of execution traces to guide the search of \textsc{Sarfgen}~\citep{sarfgen}, an APR tool for introductory programming exercises. More recently, \citet{gupta2020synthesize} in a neural program synthesis system used execution traces to repair incorrectly synthesized programs.

Details on how to download, install and run RunBugRun are provided on the project website under \url{https://github.com/giganticode/run_bug_run}. This not only includes the dataset itself, but also the infrastructure to safely execute submissions.

\printbibliography

@inproceedings{benderDangersStochasticParrots2021,
  title = {On the {{Dangers}} of {{Stochastic Parrots}}: {{Can Language Models Be Too Big}}?},
  shorttitle = {On the {{Dangers}} of {{Stochastic Parrots}}},
  booktitle = {Proceedings of the 2021 {{ACM Conference}} on {{Fairness}}, {{Accountability}}, and {{Transparency}}},
  author = {Bender, Emily M. and Gebru, Timnit and McMillan-Major, Angelina and Shmitchell, Shmargaret},
  date = {2021-03-03},
  series = {{{FAccT}} '21},
  pages = {610--623},
  publisher = {{Association for Computing Machinery}},
  location = {{New York, NY, USA}},
  doi = {10.1145/3442188.3445922},
  url = {http://doi.org/10.1145/3442188.3445922},
  urldate = {2022-08-29},
  isbn = {978-1-4503-8309-7},
}

@inproceedings{yeNeuralProgramRepair2022,
  title = {Neural Program Repair with Execution-Based Backpropagation},
  booktitle = {Proceedings of the 44th {{International Conference}} on {{Software Engineering}}},
  author = {Ye, He and Martinez, Matias and Monperrus, Martin},
  date = {2022-05-21},
  series = {{{ICSE}} '22},
  pages = {1506--1518},
  publisher = {{Association for Computing Machinery}},
  location = {{New York, NY, USA}},
  doi = {10.1145/3510003.3510222},
  url = {http://doi.org/10.1145/3510003.3510222},
  urldate = {2022-09-04},
  isbn = {978-1-4503-9221-1},
}

@article{ye2021automated,
  title={Automated patch assessment for program repair at scale},
  author={Ye, He and Martinez, Matias and Monperrus, Martin},
  journal={Empirical Software Engineering},
  volume={26},
  pages={1--38},
  year={2021},
  publisher={Springer}
}

@inproceedings{chenContractbasedProgramRepair2017,
  title = {Contract-Based Program Repair without the Contracts},
  booktitle = {2017 32nd {{IEEE}}/{{ACM International Conference}} on {{Automated Software Engineering}} ({{ASE}})},
  author = {Chen, Liushan and Pei, Yu and Furia, Carlo A.},
  date = {2017-10},
  pages = {637--647},
  doi = {10.1109/ASE.2017.8115674},
  eventtitle = {2017 32nd {{IEEE}}/{{ACM International Conference}} on {{Automated Software Engineering}} ({{ASE}})}
}

@article{yuanARJAAutomatedRepair2020,
  title = {{{ARJA}}: {{Automated Repair}} of {{Java Programs}} via {{Multi-Objective Genetic Programming}}},
  shorttitle = {{{ARJA}}},
  author = {Yuan, Yuan and Banzhaf, Wolfgang},
  date = {2020-10},
  journaltitle = {IEEE Transactions on Software Engineering},
  volume = {46},
  number = {10},
  pages = {1040--1067},
  issn = {1939-3520},
  doi = {10.1109/TSE.2018.2874648},
  eventtitle = {{{IEEE Transactions}} on {{Software Engineering}}},
}

@inproceedings{huaSketchFixToolAutomated2018,
  title = {{{SketchFix}}: A Tool for Automated Program Repair Approach Using Lazy Candidate Generation},
  shorttitle = {{{SketchFix}}},
  booktitle = {Proceedings of the 2018 26th {{ACM Joint Meeting}} on {{European Software Engineering Conference}} and {{Symposium}} on the {{Foundations}} of {{Software Engineering}}},
  author = {Hua, Jinru and Zhang, Mengshi and Wang, Kaiyuan and Khurshid, Sarfraz},
  date = {2018-10-26},
  series = {{{ESEC}}/{{FSE}} 2018},
  pages = {888--891},
  publisher = {{Association for Computing Machinery}},
  location = {{New York, NY, USA}},
  doi = {10.1145/3236024.3264600},
  url = {http://doi.org/10.1145/3236024.3264600},
  urldate = {2021-09-03},
  isbn = {978-1-4503-5573-5},
}

@misc{yeSelfAPRSelfsupervisedProgram2022,
  title = {{{SelfAPR}}: {{Self-supervised Program Repair}} with {{Test Execution Diagnostics}}},
  shorttitle = {{{SelfAPR}}},
  author = {Ye, He and Martinez, Matias and Luo, Xiapu and Zhang, Tao and Monperrus, Martin},
  date = {2022-04-04},
  number = {arXiv:2203.12755},
  eprint = {2203.12755},
  eprinttype = {arxiv},
  primaryclass = {cs},
  publisher = {{arXiv}},
  doi = {10.48550/arXiv.2203.12755},
  url = {http://arxiv.org/abs/2203.12755},
  urldate = {2022-09-04},
  archiveprefix = {arXiv},
}

@inproceedings{justDefects4JDatabaseExisting2014,
  title = {{{Defects4J}}: A Database of Existing Faults to Enable Controlled Testing Studies for {{Java}} Programs},
  shorttitle = {{{Defects4J}}},
  booktitle = {Proceedings of the 2014 {{International Symposium}} on {{Software Testing}} and {{Analysis}}},
  author = {Just, René and Jalali, Darioush and Ernst, Michael D.},
  date = {2014-07-21},
  series = {{{ISSTA}} 2014},
  pages = {437--440},
  publisher = {{Association for Computing Machinery}},
  location = {{New York, NY, USA}},
  doi = {10.1145/2610384.2628055},
  url = {http://doi.org/10.1145/2610384.2628055},
  urldate = {2021-04-08},
  isbn = {978-1-4503-2645-2},
}

@inproceedings{sahaBugsJarLargeScale2018,
  title = {Bugs.Jar: {{A Large-Scale}}, {{Diverse Dataset}} of {{Real-World Java Bugs}}},
  shorttitle = {Bugs.Jar},
  booktitle = {2018 {{IEEE}}/{{ACM}} 15th {{International Conference}} on {{Mining Software Repositories}} ({{MSR}})},
  author = {Saha, Ripon and Lyu, Yingjun and Lam, Wing and Yoshida, Hiroaki and Prasad, Mukul},
  date = {2018-05},
  pages = {10--13},
  issn = {2574-3864},
  eventtitle = {2018 {{IEEE}}/{{ACM}} 15th {{International Conference}} on {{Mining Software Repositories}} ({{MSR}})},
}

@article{madeiralBearsExtensibleJava2019,
  title = {Bears: {{An Extensible Java Bug Benchmark}} for {{Automatic Program Repair Studies}}},
  shorttitle = {Bears},
  author = {Madeiral, Fernanda and Urli, Simon and Maia, Marcelo and Monperrus, Martin},
  date = {2019-02},
  journaltitle = {2019 IEEE 26th International Conference on Software Analysis, Evolution and Reengineering (SANER)},
  eprint = {1901.06024},
  eprinttype = {arxiv},
  pages = {468--478},
  doi = {10.1109/SANER.2019.8667991},
  url = {http://arxiv.org/abs/1901.06024},
  urldate = {2021-09-18},
  archiveprefix = {arXiv},
}

@inproceedings{linQuixBugsMultilingualProgram2017,
  title = {{{QuixBugs}}: A Multi-Lingual Program Repair Benchmark Set Based on the Quixey Challenge},
  shorttitle = {{{QuixBugs}}},
  booktitle = {Proceedings {{Companion}} of the 2017 {{ACM SIGPLAN International Conference}} on {{Systems}}, {{Programming}}, {{Languages}}, and {{Applications}}: {{Software}} for {{Humanity}}},
  author = {Lin, Derrick and Koppel, James and Chen, Angela and Solar-Lezama, Armando},
  date = {2017-10-22},
  series = {{{SPLASH Companion}} 2017},
  pages = {55--56},
  publisher = {{Association for Computing Machinery}},
  location = {{New York, NY, USA}},
  doi = {10.1145/3135932.3135941},
  url = {http://doi.org/10.1145/3135932.3135941},
  urldate = {2021-04-21},
  isbn = {978-1-4503-5514-8},
}

@inproceedings{tanCodeflawsProgrammingCompetition2017,
  title = {Codeflaws: A Programming Competition Benchmark for Evaluating Automated Program Repair Tools},
  shorttitle = {Codeflaws},
  booktitle = {2017 {{IEEE}}/{{ACM}} 39th {{International Conference}} on {{Software Engineering Companion}} ({{ICSE-C}})},
  author = {Tan, Shin Hwei and Yi, Jooyong and {Yulis} and Mechtaev, Sergey and Roychoudhury, Abhik},
  date = {2017-05},
  pages = {180--182},
  doi = {10.1109/ICSE-C.2017.76},
  eventtitle = {2017 {{IEEE}}/{{ACM}} 39th {{International Conference}} on {{Software Engineering Companion}} ({{ICSE-C}})},
}

@article{legouesManyBugsIntroClassBenchmarks2015,
  title = {The {{ManyBugs}} and {{IntroClass Benchmarks}} for {{Automated Repair}} of {{C Programs}}},
  author = {Le Goues, Claire and Holtschulte, Neal and Smith, Edward K. and Brun, Yuriy and Devanbu, Premkumar and Forrest, Stephanie and Weimer, Westley},
  date = {2015-12},
  journaltitle = {IEEE Transactions on Software Engineering},
  volume = {41},
  number = {12},
  pages = {1236--1256},
  issn = {1939-3520},
  doi = {10.1109/TSE.2015.2454513},
  eventtitle = {{{IEEE Transactions}} on {{Software Engineering}}},
}

@report{durieuxIntroClassJavaBenchmark2972016a,
  type = {Research Report},
  title = {{{IntroClassJava}}: {{A Benchmark}} of 297 {{Small}} and {{Buggy Java Programs}}},
  shorttitle = {{{IntroClassJava}}},
  author = {Durieux, Thomas and Monperrus, Martin},
  date = {2016},
  number = {hal-01272126},
  institution = {{Universite Lille 1}},
  url = {https://hal.archives-ouvertes.fr/hal-01272126},
  urldate = {2021-09-18},
}

@inproceedings{tomassiBugSwarmMiningContinuously2019,
  title = {{{BugSwarm}}: Mining and Continuously Growing a Dataset of Reproducible Failures and Fixes},
  shorttitle = {{{BugSwarm}}},
  booktitle = {Proceedings of the 41st {{International Conference}} on {{Software Engineering}}},
  author = {Tomassi, David A. and Dmeiri, Naji and Wang, Yichen and Bhowmick, Antara and Liu, Yen-Chuan and Devanbu, Premkumar T. and Vasilescu, Bogdan and Rubio-González, Cindy},
  date = {2019-05-25},
  series = {{{ICSE}} '19},
  pages = {339--349},
  publisher = {{IEEE Press}},
  location = {{Montreal, Quebec, Canada}},
  doi = {10.1109/ICSE.2019.00048},
  url = {http://doi.org/10.1109/ICSE.2019.00048},
  urldate = {2021-09-20},
}

@inproceedings{widyasariBugsInPyDatabaseExisting2020,
  title = {{{BugsInPy}}: A Database of Existing Bugs in {{Python}} Programs to Enable Controlled Testing and Debugging Studies},
  shorttitle = {{{BugsInPy}}},
  booktitle = {Proceedings of the 28th {{ACM Joint Meeting}} on {{European Software Engineering Conference}} and {{Symposium}} on the {{Foundations}} of {{Software Engineering}}},
  author = {Widyasari, Ratnadira and Sim, Sheng Qin and Lok, Camellia and Qi, Haodi and Phan, Jack and Tay, Qijin and Tan, Constance and Wee, Fiona and Tan, Jodie Ethelda and Yieh, Yuheng and Goh, Brian and Thung, Ferdian and Kang, Hong Jin and Hoang, Thong and Lo, David and Ouh, Eng Lieh},
  date = {2020-11-08},
  series = {{{ESEC}}/{{FSE}} 2020},
  pages = {1556--1560},
  publisher = {{Association for Computing Machinery}},
  location = {{New York, NY, USA}},
  doi = {10.1145/3368089.3417943},
  url = {http://doi.org/10.1145/3368089.3417943},
  urldate = {2021-09-20},
  isbn = {978-1-4503-7043-1},
}

@inproceedings{bentonDefextsCuratedDataset2019,
  title = {Defexts: {{A Curated Dataset}} of {{Reproducible Real-World Bugs}} for {{Modern JVM Languages}}},
  shorttitle = {Defexts},
  booktitle = {2019 {{IEEE}}/{{ACM}} 41st {{International Conference}} on {{Software Engineering}}: {{Companion Proceedings}} ({{ICSE-Companion}})},
  author = {Benton, Samuel and Ghanbari, Ali and Zhang, Lingming},
  date = {2019-05},
  pages = {47--50},
  issn = {2574-1934},
  doi = {10.1109/ICSE-Companion.2019.00035},
  eventtitle = {2019 {{IEEE}}/{{ACM}} 41st {{International Conference}} on {{Software Engineering}}: {{Companion Proceedings}} ({{ICSE-Companion}})},
}

@inproceedings{gyimesiBugsJSBenchmarkJavaScript2019,
  title = {{{BugsJS}}: A {{Benchmark}} of {{JavaScript Bugs}}},
  shorttitle = {{{BugsJS}}},
  booktitle = {2019 12th {{IEEE Conference}} on {{Software Testing}}, {{Validation}} and {{Verification}} ({{ICST}})},
  author = {Gyimesi, Péter and Vancsics, Béla and Stocco, Andrea and Mazinanian, Davood and Beszédes, Árpád and Ferenc, Rudolf and Mesbah, Ali},
  date = {2019-04},
  pages = {90--101},
  issn = {2159-4848},
  doi = {10.1109/ICST.2019.00019},
  eventtitle = {2019 12th {{IEEE Conference}} on {{Software Testing}}, {{Validation}} and {{Verification}} ({{ICST}})},
}

@inproceedings{karampatsisHowOftenSingleStatement2020,
  title = {How {{Often Do Single-Statement Bugs Occur}}? {{The ManySStuBs4J Dataset}}},
  shorttitle = {How {{Often Do Single-Statement Bugs Occur}}?},
  booktitle = {Proceedings of the 17th {{International Conference}} on {{Mining Software Repositories}}},
  author = {Karampatsis, Rafael-Michael and Sutton, Charles},
  date = {2020-06-29},
  series = {{{MSR}} '20},
  pages = {573--577},
  publisher = {{Association for Computing Machinery}},
  location = {{New York, NY, USA}},
  doi = {10.1145/3379597.3387491},
  url = {https://doi.org/10.1145/3379597.3387491},
  urldate = {2021-04-14},
  isbn = {978-1-4503-7517-7},
}

@unpublished{chenCodRepMachineLearning2018,
  title = {The {{CodRep Machine Learning}} on {{Source Code Competition}}},
  author = {Chen, Zimin and Monperrus, Martin},
  date = {2018-11-14},
  eprint = {1807.03200},
  eprinttype = {arxiv},
  primaryclass = {cs},
  url = {http://arxiv.org/abs/1807.03200},
  urldate = {2021-03-04},
  archiveprefix = {arXiv},
}

@inproceedings{tufanoEmpiricalInvestigationLearning2018,
  title = {An {{Empirical Investigation}} into {{Learning Bug-Fixing Patches}} in the {{Wild}} via {{Neural Machine Translation}}},
  booktitle = {2018 33rd {{IEEE}}/{{ACM International Conference}} on {{Automated Software Engineering}} ({{ASE}})},
  author = {Tufano, M. and Watson, C. and Bavota, G. and di Penta, M. and White, M. and Poshyvanyk, D.},
  date = {2018-09},
  pages = {832--837},
  issn = {2643-1572},
  doi = {10.1145/3238147.3240732},
  eventtitle = {2018 33rd {{IEEE}}/{{ACM International Conference}} on {{Automated Software Engineering}} ({{ASE}})},
}

@unpublished{tufanoEmpiricalStudyLearning2019,
  title = {An {{Empirical Study}} on {{Learning Bug-Fixing Patches}} in the {{Wild}} via {{Neural Machine Translation}}},
  author = {Tufano, Michele and Watson, Cody and Bavota, Gabriele and Di Penta, Massimiliano and White, Martin and Poshyvanyk, Denys},
  date = {2019-05-20},
  eprint = {1812.08693},
  eprinttype = {arxiv},
  primaryclass = {cs},
  url = {http://arxiv.org/abs/1812.08693},
  urldate = {2021-03-04},
  archiveprefix = {arXiv},
}

@inproceedings{lutellierCoCoNuTCombiningContextaware2020,
  title = {{{CoCoNuT}}: Combining Context-Aware Neural Translation Models Using Ensemble for Program Repair},
  shorttitle = {{{CoCoNuT}}},
  booktitle = {Proceedings of the 29th {{ACM SIGSOFT International Symposium}} on {{Software Testing}} and {{Analysis}}},
  author = {Lutellier, Thibaud and Pham, Hung Viet and Pang, Lawrence and Li, Yitong and Wei, Moshi and Tan, Lin},
  date = {2020-07-18},
  series = {{{ISSTA}} 2020},
  pages = {101--114},
  publisher = {{Association for Computing Machinery}},
  location = {{New York, NY, USA}},
  doi = {10.1145/3395363.3397369},
  url = {http://doi.org/10.1145/3395363.3397369},
  urldate = {2021-04-01},
  isbn = {978-1-4503-8008-9},
}

@unpublished{monperrusMegadiffDataset600k2021,
  title = {Megadiff: {{A Dataset}} of 600k {{Java Source Code Changes Categorized}} by {{Diff Size}}},
  shorttitle = {Megadiff},
  author = {Monperrus, Martin and Martinez, Matias and Ye, He and Madeiral, Fernanda and Durieux, Thomas and Yu, Zhongxing},
  date = {2021-08-10},
  eprint = {2108.04631},
  eprinttype = {arxiv},
  primaryclass = {cs},
  url = {http://arxiv.org/abs/2108.04631},
  urldate = {2021-09-19},
  archiveprefix = {arXiv},
}

@inproceedings{csuvikFixJSDatasetBugfixing2022,
  title = {{{FixJS}}: {{A Dataset}} of {{Bug-fixing JavaScript Commits}}},
  shorttitle = {{{FixJS}}},
  booktitle = {2022 {{IEEE}}/{{ACM}} 19th {{International Conference}} on {{Mining Software Repositories}} ({{MSR}})},
  author = {Csuvik, Viktor and Vidács, László},
  date = {2022-05},
  pages = {712--716},
  issn = {2574-3864},
  doi = {10.1145/3524842.3528480},
  eventtitle = {2022 {{IEEE}}/{{ACM}} 19th {{International Conference}} on {{Mining Software Repositories}} ({{MSR}})},
}

@inproceedings{richterTSSB3MMiningSingle2022,
  title = {{{TSSB-3M}}: {{Mining}} Single Statement Bugs at Massive Scale},
  shorttitle = {{{TSSB-3M}}},
  booktitle = {2022 {{IEEE}}/{{ACM}} 19th {{International Conference}} on {{Mining Software Repositories}} ({{MSR}})},
  author = {Richter, Cedric and Wehrheim, Heike},
  date = {2022-05},
  pages = {418--422},
  issn = {2574-3864},
  doi = {10.1145/3524842.3528505},
  eventtitle = {2022 {{IEEE}}/{{ACM}} 19th {{International Conference}} on {{Mining Software Repositories}} ({{MSR}})},
}

@inproceedings{latozaMaintainingMentalModels2006,
  title = {Maintaining Mental Models: A Study of Developer Work Habits},
  shorttitle = {Maintaining Mental Models},
  booktitle = {Proceeding of the 28th International Conference on {{Software}} Engineering  - {{ICSE}} '06},
  author = {LaToza, Thomas D. and Venolia, Gina and DeLine, Robert},
  date = {2006},
  pages = {492},
  publisher = {{ACM Press}},
  location = {{Shanghai, China}},
  doi = {10.1145/1134285.1134355},
  url = {http://portal.acm.org/citation.cfm?doid=1134285.1134355},
  urldate = {2021-03-15},
  eventtitle = {Proceeding of the 28th International Conference},
  isbn = {978-1-59593-375-1},
  langid = {english},
}

@book{brittonReversibleDebuggingSoftware,
  title = {Reversible {{Debugging Software}} “{{Quantify}} the Time and Cost Saved Using Reversible Debuggers”},
  author = {Britton, Tom and Jeng, Lisa and Carver, Graham and Cheak, Paul},
}

@article{krasnerCostPoorSoftware2020,
  title = {The {{Cost}} of {{Poor Software Quality}} in the {{US}}: {{A}} 2020 {{Report}}},
  author = {Krasner, Herb},
  date = {2020},
  pages = {46},
  langid = {english},
}

@inproceedings{kimHowLongDid2006,
  title = {How Long Did It Take to Fix Bugs?},
  booktitle = {Proceedings of the 2006 International Workshop on {{Mining}} Software Repositories},
  author = {Kim, Sunghun and Whitehead, E. James},
  date = {2006-05-22},
  series = {{{MSR}} '06},
  pages = {173--174},
  publisher = {{Association for Computing Machinery}},
  location = {{New York, NY, USA}},
  doi = {10.1145/1137983.1138027},
  url = {http://doi.org/10.1145/1137983.1138027},
  urldate = {2021-06-03},
  isbn = {978-1-59593-397-3},
}

@article{panUnderstandingBugFix2009,
  title = {Toward an Understanding of Bug Fix Patterns},
  author = {Pan, Kai and Kim, Sunghun and Whitehead, E. James},
  date = {2009-06-01},
  journaltitle = {Empirical Software Engineering},
  shortjournal = {Empirical Softw. Engg.},
  volume = {14},
  number = {3},
  pages = {286--315},
  issn = {1382-3256},
  doi = {10.1007/s10664-008-9077-5},
  url = {http://doi.org/10.1007/s10664-008-9077-5},
  urldate = {2021-06-08},
}

@article{gouesGenProgGenericMethod2012,
  title = {{{GenProg}}: {{A Generic Method}} for {{Automatic Software Repair}}},
  shorttitle = {{{GenProg}}},
  author = {Goues, C. Le and Nguyen, T. and Forrest, S. and Weimer, W.},
  date = {2012-01},
  journaltitle = {IEEE Transactions on Software Engineering},
  volume = {38},
  number = {1},
  pages = {54--72},
  issn = {1939-3520},
  doi = {10.1109/TSE.2011.104},
  eventtitle = {{{IEEE Transactions}} on {{Software Engineering}}},
}

@inproceedings{forrestGeneticProgrammingApproach2009,
  title = {A Genetic Programming Approach to Automated Software Repair},
  booktitle = {Proceedings of the 11th {{Annual}} Conference on {{Genetic}} and Evolutionary Computation},
  author = {Forrest, Stephanie and Nguyen, ThanhVu and Weimer, Westley and Le Goues, Claire},
  date = {2009-07-08},
  series = {{{GECCO}} '09},
  pages = {947--954},
  publisher = {{Association for Computing Machinery}},
  location = {{New York, NY, USA}},
  doi = {10.1145/1569901.1570031},
  url = {http://doi.org/10.1145/1569901.1570031},
  urldate = {2021-02-22},
  isbn = {978-1-60558-325-9}
}

@inproceedings{durieuxDynaMothDynamicCode2016,
  title = {{{DynaMoth}}: {{Dynamic Code Synthesis}} for {{Automatic Program Repair}}},
  shorttitle = {{{DynaMoth}}},
  booktitle = {2016 {{IEEE}}/{{ACM}} 11th {{International Workshop}} in {{Automation}} of {{Software Test}} ({{AST}})},
  author = {Durieux, Thomas and Monperrus, Martin},
  date = {2016-05},
  pages = {85--91},
  doi = {10.1109/AST.2016.021},
  eventtitle = {2016 {{IEEE}}/{{ACM}} 11th {{International Workshop}} in {{Automation}} of {{Software Test}} ({{AST}})},
}

@inproceedings{zhong2022standup4npr,
  title={StandUp4NPR: Standardizing SetUp for Empirically Comparing Neural Program Repair Systems},
  author={Zhong, Wenkang and Ge, Hongliang and Ai, Hongfei and Li, Chuanyi and Liu, Kui and Ge, Jidong and Luo, Bin},
  booktitle={Conference on Automated Software Engineering (ASE’22)},
  year={2022}
}

@article{code4bench,
title = {Code4Bench: A multidimensional benchmark of Codeforces data for different program analysis techniques},
journal = {Journal of Computer Languages},
volume = {53},
pages = {38-52},
year = {2019},
issn = {2590-1184},
doi = {https://doi.org/10.1016/j.cola.2019.03.006},
url = {https://www.sciencedirect.com/science/article/pii/S1045926X18302489},
author = {Amirabbas Majd and Mojtaba Vahidi-Asl and Alireza Khalilian and Ahmad Baraani-Dastjerdi and Bahman Zamani},
}

@article{majdCode4BenchMultidimensionalBenchmark2019,
  title = {{{Code4Bench}}: {{A}} Multidimensional Benchmark of {{Codeforces}} Data for Different Program Analysis Techniques},
  shorttitle = {{{Code4Bench}}},
  author = {Majd, Amirabbas and Vahidi-Asl, Mojtaba and Khalilian, Alireza and Baraani-Dastjerdi, Ahmad and Zamani, Bahman},
  date = {2019-08-01},
  journaltitle = {Journal of Computer Languages},
  shortjournal = {Journal of Computer Languages},
  volume = {53},
  pages = {38--52},
  issn = {2590-1184},
  doi = {10.1016/j.cola.2019.03.006},
  url = {https://www.sciencedirect.com/science/article/pii/S1045926X18302489},
  urldate = {2022-11-23},
  langid = {english},
}

@article{liuCriticalReviewEvaluation2021,
  title = {A Critical Review on the Evaluation of Automated Program Repair Systems},
  author = {Liu, Kui and Li, Li and Koyuncu, Anil and Kim, Dongsun and Liu, Zhe and Klein, Jacques and Bissyandé, Tegawendé F.},
  date = {2021-01},
  journaltitle = {Journal of Systems and Software},
  shortjournal = {Journal of Systems and Software},
  volume = {171},
  pages = {110817},
  issn = {01641212},
  doi = {10.1016/j.jss.2020.110817},
  url = {https://linkinghub.elsevier.com/retrieve/pii/S0164121220302156},
  urldate = {2021-03-23},
  langid = {english}
}

@inproceedings{nodaExperienceReportHow2020a,
  title = {Experience {{Report}}: {{How Effective}} Is {{Automated Program Repair}} for {{Industrial Software}}?},
  shorttitle = {Experience {{Report}}},
  booktitle = {2020 {{IEEE}} 27th {{International Conference}} on {{Software Analysis}}, {{Evolution}} and {{Reengineering}} ({{SANER}})},
  author = {Noda, Kunihiro and Nemoto, Yusuke and Hotta, Keisuke and Tanida, Hideo and Kikuchi, Shinji},
  date = {2020-02},
  pages = {612--616},
  issn = {1534-5351},
  doi = {10.1109/SANER48275.2020.9054829},
  eventtitle = {2020 {{IEEE}} 27th {{International Conference}} on {{Software Analysis}}, {{Evolution}} and {{Reengineering}} ({{SANER}})},
}

@article{camposDiscoveringCommonBugfix2019,
  title = {Discovering Common Bug-Fix Patterns: {{A}} Large-Scale Observational Study},
  shorttitle = {Discovering Common Bug-Fix Patterns},
  author = {Campos, Eduardo C. and Maia, Marcelo de A.},
  date = {2019},
  journaltitle = {Journal of Software: Evolution and Process},
  volume = {31},
  number = {7},
  pages = {e2173},
  issn = {2047-7481},
  doi = {10.1002/smr.2173},
  url = {http://onlinelibrary.wiley.com/doi/abs/10.1002/smr.2173},
  urldate = {2021-05-31},
  langid = {english},
}

@misc{puriCodeNetLargeScaleAI2021,
  title = {{{CodeNet}}: {{A Large-Scale AI}} for {{Code Dataset}} for {{Learning}} a {{Diversity}} of {{Coding Tasks}}},
  shorttitle = {{{CodeNet}}},
  author = {Puri, Ruchir and Kung, David S. and Janssen, Geert and Zhang, Wei and Domeniconi, Giacomo and Zolotov, Vladimir and Dolby, Julian and Chen, Jie and Choudhury, Mihir and Decker, Lindsey and Thost, Veronika and Buratti, Luca and Pujar, Saurabh and Ramji, Shyam and Finkler, Ulrich and Malaika, Susan and Reiss, Frederick},
  date = {2021-08-29},
  number = {arXiv:2105.12655},
  eprint = {2105.12655},
  eprinttype = {arxiv},
  primaryclass = {cs},
  publisher = {{arXiv}},
  doi = {10.48550/arXiv.2105.12655},
  url = {http://arxiv.org/abs/2105.12655},
  urldate = {2022-11-16},
  archiveprefix = {arXiv},
}

@unpublished{chenSequenceRSequencetoSequenceLearning2019,
  title = {{{SequenceR}}: {{Sequence-to-Sequence Learning}} for {{End-to-End Program Repair}}},
  shorttitle = {{{SequenceR}}},
  author = {Chen, Zimin and Kommrusch, Steve and Tufano, Michele and Pouchet, Louis-Noël and Poshyvanyk, Denys and Monperrus, Martin},
  date = {2019-09-09},
  eprint = {1901.01808},
  eprinttype = {arxiv},
  primaryclass = {cs, stat},
  doi = {10.1109/TSE.2019.2940179},
  url = {http://arxiv.org/abs/1901.01808},
  urldate = {2021-02-15},
  archiveprefix = {arXiv},
}

@unpublished{jiangCURECodeAwareNeural2021,
  title = {{{CURE}}: {{Code-Aware Neural Machine Translation}} for {{Automatic Program Repair}}},
  shorttitle = {{{CURE}}},
  author = {Jiang, Nan and Lutellier, Thibaud and Tan, Lin},
  date = {2021-02-26},
  eprint = {2103.00073},
  eprinttype = {arxiv},
  primaryclass = {cs},
  url = {http://arxiv.org/abs/2103.00073},
  urldate = {2021-03-22},
  archiveprefix = {arXiv},
}

@inproceedings{liDLFixContextbasedCode2020,
  title = {{{DLFix}}: Context-Based Code Transformation Learning for Automated Program Repair},
  shorttitle = {{{DLFix}}},
  booktitle = {Proceedings of the {{ACM}}/{{IEEE}} 42nd {{International Conference}} on {{Software Engineering}}},
  author = {Li, Yi and Wang, Shaohua and Nguyen, Tien N.},
  date = {2020-06-27},
  series = {{{ICSE}} '20},
  pages = {602--614},
  publisher = {{Association for Computing Machinery}},
  location = {{New York, NY, USA}},
  doi = {10.1145/3377811.3380345},
  url = {http://doi.org/10.1145/3377811.3380345},
  urldate = {2021-04-01},
  isbn = {978-1-4503-7121-6},
}

@inproceedings{dinellaHOPPITYLEARNINGGRAPH2020,
  title = {{{HOPPITY}}: {{LEARNING GRAPH TRANSFORMATIONS TO DETECT AND FIX BUGS IN PROGRAMS}}},
  shorttitle = {{{HOPPITY}}},
  author = {Dinella, Elizabeth and Dai, Hanjun and Li, Ziyang and Naik, Mayur and Song, Le and Wang, Ke},
  date = {2020-04},
  url = {https://iclr.cc/virtual_2020/poster_SJeqs6EFvB.html},
  urldate = {2021-03-05},
  eventtitle = {Eighth {{International Conference}} on {{Learning Representations}}},
  langid = {english},
}

@inproceedings{joshiStateFateLinguistic2020,
  title = {The {{State}} and {{Fate}} of {{Linguistic Diversity}} and {{Inclusion}} in the {{NLP World}}},
  booktitle = {Proceedings of the 58th {{Annual Meeting}} of the {{Association}} for {{Computational Linguistics}}},
  author = {Joshi, Pratik and Santy, Sebastin and Budhiraja, Amar and Bali, Kalika and Choudhury, Monojit},
  date = {2020-07},
  pages = {6282--6293},
  publisher = {{Association for Computational Linguistics}},
  location = {{Online}},
  doi = {10.18653/v1/2020.acl-main.560},
  url = {https://aclanthology.org/2020.acl-main.560},
  urldate = {2022-11-17},
  eventtitle = {{{ACL}} 2020},
}

@misc{ruder2020beyondenglish,
  author = {Ruder, Sebastian},
  title = {{Why You Should Do NLP Beyond English}},
  year = {2020},
  howpublished = {\url{http://ruder.io/nlp-beyond-english}},
}

@misc{haqueFixEvalExecutionbasedEvaluation2022,
  title = {{{FixEval}}: {{Execution-based Evaluation}} of {{Program Fixes}} for {{Programming Problems}}},
  shorttitle = {{{FixEval}}},
  author = {Haque, Md Mahim Anjum and Ahmad, Wasi Uddin and Lourentzou, Ismini and Brown, Chris},
  date = {2022-09-29},
  number = {arXiv:2206.07796},
  eprint = {2206.07796},
  eprinttype = {arxiv},
  primaryclass = {cs},
  publisher = {{arXiv}},
  doi = {10.48550/arXiv.2206.07796},
  url = {http://arxiv.org/abs/2206.07796},
  urldate = {2022-11-23},
  archiveprefix = {arXiv},
}

@misc{richterCanWeLearn2022,
  title = {Can We Learn from Developer Mistakes? {{Learning}} to Localize and Repair Real Bugs from Real Bug Fixes},
  shorttitle = {Can We Learn from Developer Mistakes?},
  author = {Richter, Cedric and Wehrheim, Heike},
  date = {2022-07-01},
  number = {arXiv:2207.00301},
  eprint = {2207.00301},
  eprinttype = {arxiv},
  primaryclass = {cs},
  publisher = {{arXiv}},
  doi = {10.48550/arXiv.2207.00301},
  url = {http://arxiv.org/abs/2207.00301},
  urldate = {2022-08-25},
  archiveprefix = {arXiv},
}

@inproceedings{allamanisSelfSupervisedBugDetection2021a,
  title = {Self-{{Supervised Bug Detection}} and {{Repair}}},
  booktitle = {Advances in {{Neural Information Processing Systems}}},
  author = {Allamanis, Miltiadis and Jackson-Flux, Henry and Brockschmidt, Marc},
  date = {2021},
  volume = {34},
  pages = {27865--27876},
  publisher = {{Curran Associates, Inc.}},
  url = {https://proceedings.neurips.cc/paper/2021/hash/ea96efc03b9a050d895110db8c4af057-Abstract.html},
  urldate = {2022-11-23},
}

@article{husain2019codesearchnet,
  title={{CodeSearchNet} challenge: Evaluating the state of semantic code search},
  author={Husain, Hamel and Wu, Ho-Hsiang and Gazit, Tiferet and Allamanis, Miltiadis and Brockschmidt, Marc},
  journal={arXiv preprint arXiv:1909.09436},
  year={2019}
}

@article{yangWhereWereRepair2021,
  title = {Where Were the Repair Ingredients for {{Defects4j}} Bugs?},
  author = {Yang, Deheng and Liu, Kui and Kim, Dongsun and Koyuncu, Anil and Kim, Kisub and Tian, Haoye and Lei, Yan and Mao, Xiaoguang and Klein, Jacques and Bissyandé, Tegawendé F.},
  date = {2021-09-10},
  journaltitle = {Empirical Software Engineering},
  shortjournal = {Empir Software Eng},
  volume = {26},
  number = {6},
  pages = {122},
  issn = {1573-7616},
  doi = {10.1007/s10664-021-10003-7},
  url = {https://doi.org/10.1007/s10664-021-10003-7},
  urldate = {2022-11-26},
  langid = {english},
}

@misc{alphacode,
  doi = {10.48550/ARXIV.2203.07814},
  url = {https://arxiv.org/abs/2203.07814},
  author = {Li, Yujia and Choi, David and Chung, Junyoung and Kushman, Nate and Schrittwieser, Julian and Leblond, Rémi and Eccles, Tom and Keeling, James and Gimeno, Felix and Lago, Agustin Dal and Hubert, Thomas and Choy, Peter and d'Autume, Cyprien de Masson and Babuschkin, Igor and Chen, Xinyun and Huang, Po-Sen and Welbl, Johannes and Gowal, Sven and Cherepanov, Alexey and Molloy, James and Mankowitz, Daniel J. and Robson, Esme Sutherland and Kohli, Pushmeet and de Freitas, Nando and Kavukcuoglu, Koray and Vinyals, Oriol},
  title = {Competition-Level Code Generation with AlphaCode},
  publisher = {arXiv},
  year = {2022},
  copyright = {Creative Commons Attribution 4.0 International}
}

@InProceedings{cardumen,
author="Martinez, Matias
and Monperrus, Martin",
editor="Colanzi, Thelma Elita
and McMinn, Phil",
title="Ultra-Large Repair Search Space with Automatically Mined Templates: The Cardumen Mode of Astor",
booktitle="Search-Based Software Engineering",
year="2018",
publisher="Springer International Publishing",
address="Cham",
pages="65--86",
isbn="978-3-319-99241-9"
}

@misc{codet5,
  doi = {codet5},
  url = {https://arxiv.org/abs/2109.00859},
  author = {Wang, Yue and Wang, Weishi and Joty, Shafiq and Hoi, Steven C. H.},
  title = {CodeT5: Identifier-aware Unified Pre-trained Encoder-Decoder Models for Code Understanding and Generation},
  publisher = {arXiv},
  year = {2021},
  copyright = {Creative Commons Attribution Non Commercial Share Alike 4.0 International}
}

@article{YE2021110825,
title = {A comprehensive study of automatic program repair on the QuixBugs benchmark},
journal = {Journal of Systems and Software},
volume = {171},
pages = {110825},
year = {2021},
issn = {0164-1212},
doi = {https://doi.org/10.1016/j.jss.2020.110825},
url = {https://www.sciencedirect.com/science/article/pii/S0164121220302193},
author = {He Ye and Matias Martinez and Thomas Durieux and Martin Monperrus},
}

@misc{huggingface,
  doi = {10.48550/ARXIV.1910.03771},
  url = {https://arxiv.org/abs/1910.03771},
  author = {Wolf, Thomas and Debut, Lysandre and Sanh, Victor and Chaumond, Julien and Delangue, Clement and Moi, Anthony and Cistac, Pierric and Rault, Tim and Louf, Rémi and Funtowicz, Morgan and Davison, Joe and Shleifer, Sam and von Platen, Patrick and Ma, Clara and Jernite, Yacine and Plu, Julien and Xu, Canwen and Scao, Teven Le and Gugger, Sylvain and Drame, Mariama and Lhoest, Quentin and Rush, Alexander M.},
  title = {HuggingFace's Transformers: State-of-the-art Natural Language Processing},
  publisher = {arXiv},
  year = {2019},
  copyright = {arXiv.org perpetual, non-exclusive license}
}

@misc{plbart,
  doi = {10.48550/ARXIV.2103.06333},
  url = {https://arxiv.org/abs/2103.06333},
  author = {Ahmad, Wasi Uddin and Chakraborty, Saikat and Ray, Baishakhi and Chang, Kai-Wei},
  title = {Unified Pre-training for Program Understanding and Generation},
  publisher = {arXiv},
  year = {2021},
  copyright = {arXiv.org perpetual, non-exclusive license}
}

@misc{flacoco,
  doi = {10.48550/ARXIV.2111.12513},
  url = {https://arxiv.org/abs/2111.12513},
  author = {Silva, André and Martinez, Matias and Danglot, Benjamin and Ginelli, Davide and Monperrus, Martin},
  title = {FLACOCO: Fault Localization for Java based on Industry-grade Coverage},
  publisher = {arXiv},
  year = {2021},
  copyright = {Creative Commons Attribution Share Alike 4.0 International}
}

@inproceedings{campos2012gzoltar,
  title={Gzoltar: an eclipse plug-in for testing and debugging},
  author={Campos, Jos{\'e} and Riboira, Andr{\'e} and Perez, Alexandre and Abreu, Rui},
  booktitle={Proceedings of the 27th IEEE/ACM international conference on automated software engineering},
  pages={378--381},
  year={2012}
}

@inproceedings{martinez2016astor,
  title={Astor: A program repair library for java},
  author={Martinez, Matias and Monperrus, Martin},
  booktitle={Proceedings of the 25th International Symposium on Software Testing and Analysis},
  pages={441--444},
  year={2016}
}

@inproceedings{smithCureWorseDisease2015,
	location = {New York, {NY}, {USA}},
	title = {Is the cure worse than the disease? overfitting in automated program repair},
	isbn = {978-1-4503-3675-8},
	url = {http://doi.org/10.1145/2786805.2786825},
	doi = {10.1145/2786805.2786825},
	series = {{ESEC}/{FSE} 2015},
	shorttitle = {Is the cure worse than the disease?},
	pages = {532--543},
	booktitle = {Proceedings of the 2015 10th Joint Meeting on Foundations of Software Engineering},
	publisher = {Association for Computing Machinery},
	author = {Smith, Edward K. and Barr, Earl T. and Le Goues, Claire and Brun, Yuriy},
	urldate = {2021-02-18},
	date = {2015-08-30},
}

@article{wang2017dynamic,
  title={Dynamic neural program embedding for program repair},
  author={Wang, Ke and Singh, Rishabh and Su, Zhendong},
  journal={arXiv preprint arXiv:1711.07163},
  year={2017}
}

@article{gupta2020synthesize,
  title={Synthesize, execute and debug: Learning to repair for neural program synthesis},
  author={Gupta, Kavi and Christensen, Peter Ebert and Chen, Xinyun and Song, Dawn},
  journal={Advances in Neural Information Processing Systems},
  volume={33},
  pages={17685--17695},
  year={2020}
}

@article{drainDeepDebugFixingPython2021,
	title = {{DeepDebug}: Fixing Python Bugs Using Stack Traces, Backtranslation, and Code Skeletons},
	url = {http://arxiv.org/abs/2105.09352},
	shorttitle = {{DeepDebug}},
	journaltitle = {{arXiv}:2105.09352 [cs]},
	author = {Drain, Dawn and Clement, Colin B. and Serrato, Guillermo and Sundaresan, Neel},
	urldate = {2021-09-17},
	date = {2021-05-19},
	eprinttype = {arxiv},
	eprint = {2105.09352},
}

@inproceedings{qiAnalysisPatchPlausibility2015,
	location = {New York, {NY}, {USA}},
	title = {An analysis of patch plausibility and correctness for generate-and-validate patch generation systems},
	isbn = {978-1-4503-3620-8},
	url = {http://doi.org/10.1145/2771783.2771791},
	doi = {10.1145/2771783.2771791},
	series = {{ISSTA} 2015},
	pages = {24--36},
	booktitle = {Proceedings of the 2015 International Symposium on Software Testing and Analysis},
	publisher = {Association for Computing Machinery},
	author = {Qi, Zichao and Long, Fan and Achour, Sara and Rinard, Martin},
	urldate = {2021-03-31},
	date = {2015-07-13},
}

@article{coelho2017exception,
  title={Exception handling bug hazards in Android},
  author={Coelho, Roberta and Almeida, Lucas and Gousios, Georgios and Deursen, Arie van and Treude, Christoph},
  journal={Empirical Software Engineering},
  volume={22},
  number={3},
  pages={1264--1304},
  year={2017},
  publisher={Springer}
}

@INPROCEEDINGS{comprehensivequixbugs,
  author={Ye, He and Martinez, Matias and Durieux, Thomas and Monperrus, Martin},
  booktitle={2019 IEEE 1st International Workshop on Intelligent Bug Fixing (IBF)}, 
  title={A Comprehensive Study of Automatic Program Repair on the QuixBugs Benchmark}, 
  year={2019},
  volume={},
  number={},
  pages={1-10},
  doi={10.1109/IBF.2019.8665475}}

@inproceedings{plasticsurgery,
author = {Barr, Earl T. and Brun, Yuriy and Devanbu, Premkumar and Harman, Mark and Sarro, Federica},
title = {The Plastic Surgery Hypothesis},
year = {2014},
isbn = {9781450330565},
publisher = {Association for Computing Machinery},
address = {New York, NY, USA},
url = {https://doi-org.libproxy.unibz.it/10.1145/2635868.2635898},
doi = {10.1145/2635868.2635898},
booktitle = {Proceedings of the 22nd ACM SIGSOFT International Symposium on Foundations of Software Engineering},
pages = {306–317},
numpages = {12},
location = {Hong Kong, China},
series = {FSE 2014}
}

@inproceedings{bennett2022some,
  title={Some automatically generated patches are more likely to be correct than others: an analysis of Defects4J patch features},
  author={Bennett, Gareth and Hall, Tracy and Bowes, David},
  booktitle={Proceedings of the Third International Workshop on Automated Program Repair},
  pages={46--52},
  year={2022}
}

@inproceedings{falleri2014fine,
  title={Fine-grained and accurate source code differencing},
  author={Falleri, Jean-R{\'e}my and Morandat, Flor{\'e}al and Blanc, Xavier and Martinez, Matias and Monperrus, Martin},
  booktitle={Proceedings of the 29th ACM/IEEE international conference on Automated software engineering},
  pages={313--324},
  year={2014}
}

@inproceedings{le2019reliability,
  title={On reliability of patch correctness assessment},
  author={Le, Xuan-Bach D and Bao, Lingfeng and Lo, David and Xia, Xin and Li, Shanping and Pasareanu, Corina},
  booktitle={2019 IEEE/ACM 41st International Conference on Software Engineering (ICSE)},
  pages={524--535},
  year={2019},
  organization={IEEE}
}

@misc{ney2021showYourWork,
  doi = {10.48550/ARXIV.2112.00114},
  url = {https://arxiv.org/abs/2112.00114},
  author = {Nye, Maxwell and Andreassen, Anders Johan and Gur-Ari, Guy and Michalewski, Henryk and Austin, Jacob and Bieber, David and Dohan, David and Lewkowycz, Aitor and Bosma, Maarten and Luan, David and Sutton, Charles and Odena, Augustus},
  title = {Show Your Work: Scratchpads for Intermediate Computation with Language Models},
  publisher = {arXiv},
  year = {2021},
  copyright = {arXiv.org perpetual, non-exclusive license}
}

@article{schick2023toolformer,
  title={Toolformer: Language Models Can Teach Themselves to Use Tools},
  author={Schick, Timo and Dwivedi-Yu, Jane and Dessi, Roberto and Raileanu, Roberta and Lomeli, Maria and Zettlemoyer, Luke and Cancedda, Nicola and Scialom, Thomas},
  journal={arXiv preprint arXiv:2302.04761},
  year={2023}
}

@inproceedings{herzigImpactTangledCode2013,
  title = {The Impact of Tangled Code Changes},
  booktitle = {2013 10th {{Working Conference}} on {{Mining Software Repositories}} ({{MSR}})},
  author = {Herzig, Kim and Zeller, Andreas},
  date = {2013-05},
  pages = {121--130},
  issn = {2160-1860},
  doi = {10.1109/MSR.2013.6624018},
  eventtitle = {2013 10th {{Working Conference}} on {{Mining Software Repositories}} ({{MSR}})},
}

@inproceedings{babiiMiningSoftwareRepositories2021,
  title = {Mining {{Software Repositories}} with a {{Collaborative Heuristic Repository}}},
  booktitle = {2021 {{IErEE}}/{{ACM}} 43rd {{International Conference}} on {{Software Engineering}}: {{New Ideas}} and {{Emerging Results}} ({{ICSE-NIER}})},
  author = {Babii, Hlib and Prenner, Julian Aron and Stricker, Laurin and Karmakar, Anjan and Janes, Andrea and Robbes, Romain},
  date = {2021-05},
  pages = {106--110},
  doi = {10.1109/ICSE-NIER52604.2021.00030},
  eventtitle = {2021 {{IEEE}}/{{ACM}} 43rd {{International Conference}} on {{Software Engineering}}: {{New Ideas}} and {{Emerging Results}} ({{ICSE-NIER}})},
}

@inproceedings{xiong2018identifying,
  title={Identifying patch correctness in test-based program repair},
  author={Xiong, Yingfei and Liu, Xinyuan and Zeng, Muhan and Zhang, Lu and Huang, Gang},
  booktitle={Proceedings of the 40th international conference on software engineering},
  pages={789--799},
  year={2018}
}

@inproceedings{xin2017identifying,
  title={Identifying test-suite-overfitted patches through test case generation},
  author={Xin, Qi and Reiss, Steven P},
  booktitle={Proceedings of the 26th ACM SIGSOFT international symposium on software testing and analysis},
  pages={226--236},
  year={2017}
}

@article{sarfgen,
author = {Wang, Ke and Singh, Rishabh and Su, Zhendong},
title = {Search, Align, and Repair: Data-Driven Feedback Generation for Introductory Programming Exercises},
year = {2018},
issue_date = {April 2018},
publisher = {Association for Computing Machinery},
address = {New York, NY, USA},
volume = {53},
number = {4},
issn = {0362-1340},
url = {https://doi-org.libproxy.unibz.it/10.1145/3296979.3192384},
doi = {10.1145/3296979.3192384},
journal = {SIGPLAN Not.},
month = {jun},
pages = {481–495},
numpages = {15}
}

@inproceedings{zhong2015empirical,
	title={An empirical study on real bug fixes},
	author={Zhong, Hao and Su, Zhendong},
	booktitle={2015 IEEE/ACM 37th IEEE International Conference on Software Engineering},
	volume={1},
	pages={913--923},
	year={2015},
	organization={IEEE}
}

@inproceedings{saha2017elixir,
	title={Elixir: Effective object-oriented program repair},
	author={Saha, Ripon K and Lyu, Yingjun and Yoshida, Hiroaki and Prasad, Mukul R},
	booktitle={2017 32nd IEEE/ACM International Conference on Automated Software Engineering (ASE)},
	pages={648--659},
	year={2017},
	organization={IEEE}
}

@inproceedings{durieux2019empirical,
	title={Empirical review of Java program repair tools: A large-scale experiment on 2,141 bugs and 23,551 repair attempts},
	author={Durieux, Thomas and Madeiral, Fernanda and Martinez, Matias and Abreu, Rui},
	booktitle={Proceedings of the 2019 27th ACM Joint Meeting on European Software Engineering Conference and Symposium on the Foundations of Software Engineering},
	pages={302--313},
	year={2019}
}

@misc{codebleu,
	doi = {10.48550/ARXIV.2009.10297},
	
	url = {https://arxiv.org/abs/2009.10297},
	
	author = {Ren, Shuo and Guo, Daya and Lu, Shuai and Zhou, Long and Liu, Shujie and Tang, Duyu and Sundaresan, Neel and Zhou, Ming and Blanco, Ambrosio and Ma, Shuai},
	
	title = {CodeBLEU: a Method for Automatic Evaluation of Code Synthesis},
	
	publisher = {arXiv},
	
	year = {2020},
	
	copyright = {arXiv.org perpetual, non-exclusive license}
}

@INPROCEEDINGS{flaky,
	
	author={Qin, Yihao and Wang, Shangwen and Liu, Kui and Mao, Xiaoguang and Bissyandé, Tegawendé F.},
	
	booktitle={2021 IEEE International Conference on Software Analysis, Evolution and Reengineering (SANER)}, 
	
	title={On the Impact of Flaky Tests in Automated Program Repair}, 
	
	year={2021},
	
	volume={},
	
	number={},
	
	pages={295-306},
	
	doi={10.1109/SANER50967.2021.00035}}

@inproceedings{zoph-etal-2016-transfer,
	title = "Transfer Learning for Low-Resource Neural Machine Translation",
	author = "Zoph, Barret  and
	Yuret, Deniz  and
	May, Jonathan  and
	Knight, Kevin",
	booktitle = "Proceedings of the 2016 Conference on Empirical Methods in Natural Language Processing",
	month = nov,
	year = "2016",
	address = "Austin, Texas",
	publisher = "Association for Computational Linguistics",
	url = "https://aclanthology.org/D16-1163",
	doi = "10.18653/v1/D16-1163",
	pages = "1568--1575",
}

@inproceedings{multilingual-training,
	author = {Ahmed, Toufique and Devanbu, Premkumar},
	title = {Multilingual Training for Software Engineering},
	year = {2022},
	isbn = {9781450392211},
	publisher = {Association for Computing Machinery},
	address = {New York, NY, USA},
	url = {https://doi-org.libproxy.unibz.it/10.1145/3510003.3510049},
	doi = {10.1145/3510003.3510049},
	booktitle = {Proceedings of the 44th International Conference on Software Engineering},
	pages = {1443–1455},
	numpages = {13},
	location = {Pittsburgh, Pennsylvania},
	series = {ICSE '22}
}

@inproceedings{bug-detector-dist-shift,
	title = 	 {On Distribution Shift in Learning-based Bug Detectors},
	author =       {He, Jingxuan and Beurer-Kellner, Luca and Vechev, Martin},
	booktitle = 	 {Proceedings of the 39th International Conference on Machine Learning},
	pages = 	 {8559--8580},
	year = 	 {2022},
	editor = 	 {Chaudhuri, Kamalika and Jegelka, Stefanie and Song, Le and Szepesvari, Csaba and Niu, Gang and Sabato, Sivan},
	volume = 	 {162},
	series = 	 {Proceedings of Machine Learning Research},
	month = 	 {17--23 Jul},
	publisher =    {PMLR},
	url = 	 {https://proceedings.mlr.press/v162/he22a.html}
}

@ARTICLE{arjanmt,
  author={Li, Dongcheng and Wong, W. Eric and Jian, Mingyong and Geng, Yi and Chau, Matthew},
  journal={IEEE Access}, 
  title={Improving Search-Based Automatic Program Repair With Neural Machine Translation}, 
  year={2022},
  volume={10},
  number={},
  pages={51167-51175},
  doi={10.1109/ACCESS.2022.3164780}}

@misc{coderl,
  doi = {10.48550/ARXIV.2207.01780},
  url = {https://arxiv.org/abs/2207.01780},
  author = {Le, Hung and Wang, Yue and Gotmare, Akhilesh Deepak and Savarese, Silvio and Hoi, Steven C. H.},
  title = {CodeRL: Mastering Code Generation through Pretrained Models and Deep Reinforcement Learning},
  publisher = {arXiv},
  year = {2022},
  copyright = {arXiv.org perpetual, non-exclusive license}
}

\appendix
\section{Further Tables and Figures}

\begin{figure}[htb]
	\begin{center}
		\includegraphics[width=\linewidth]{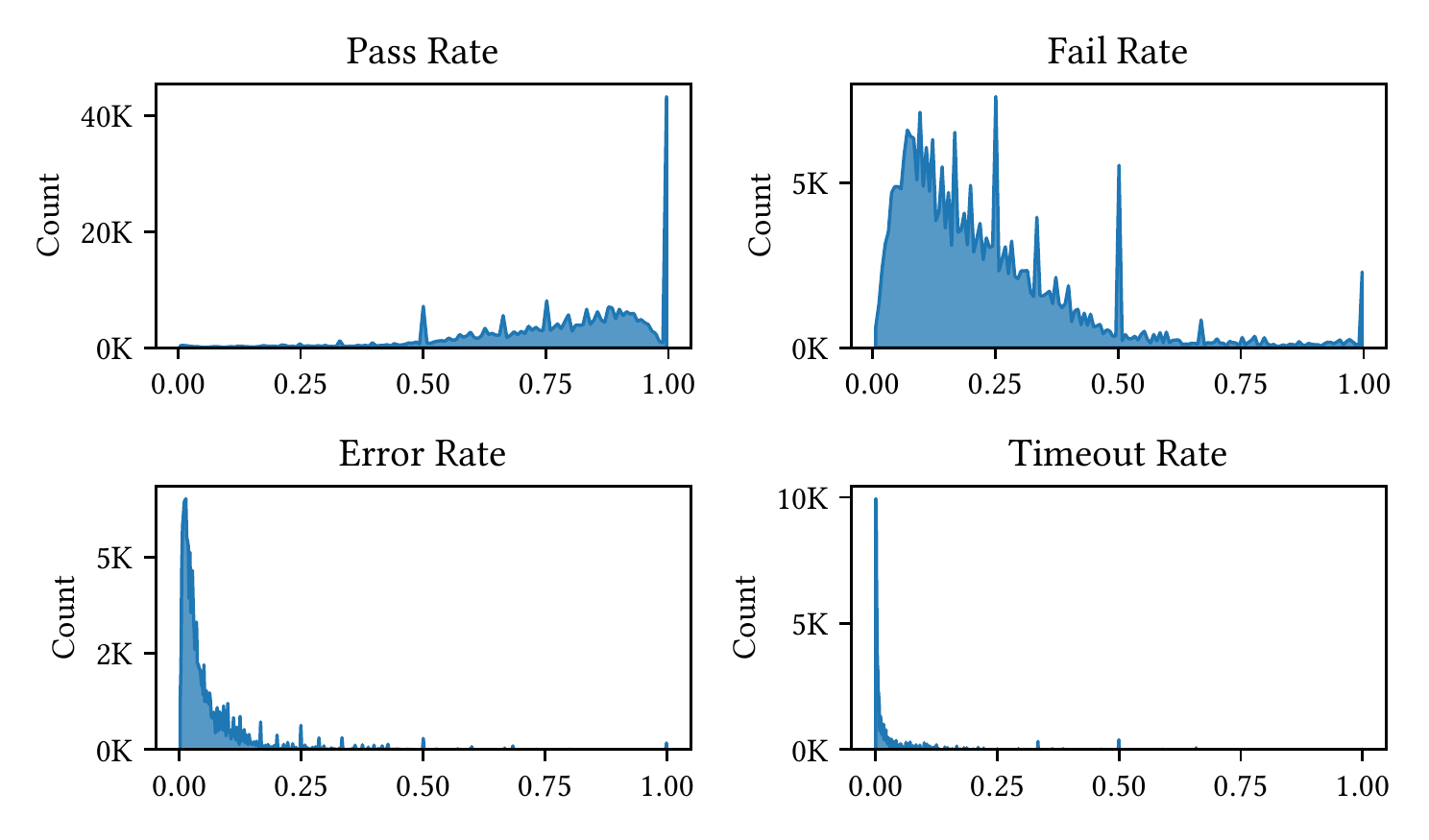}
	\end{center}
	\caption{Distribution of test cases over pass, fail, error and timeout rates (\emph{before} test case filtering). }
	\label{fig:test-rates}
\end{figure}

\begin{figure}[htb]
	\begin{center}
		\includegraphics[width=\linewidth]{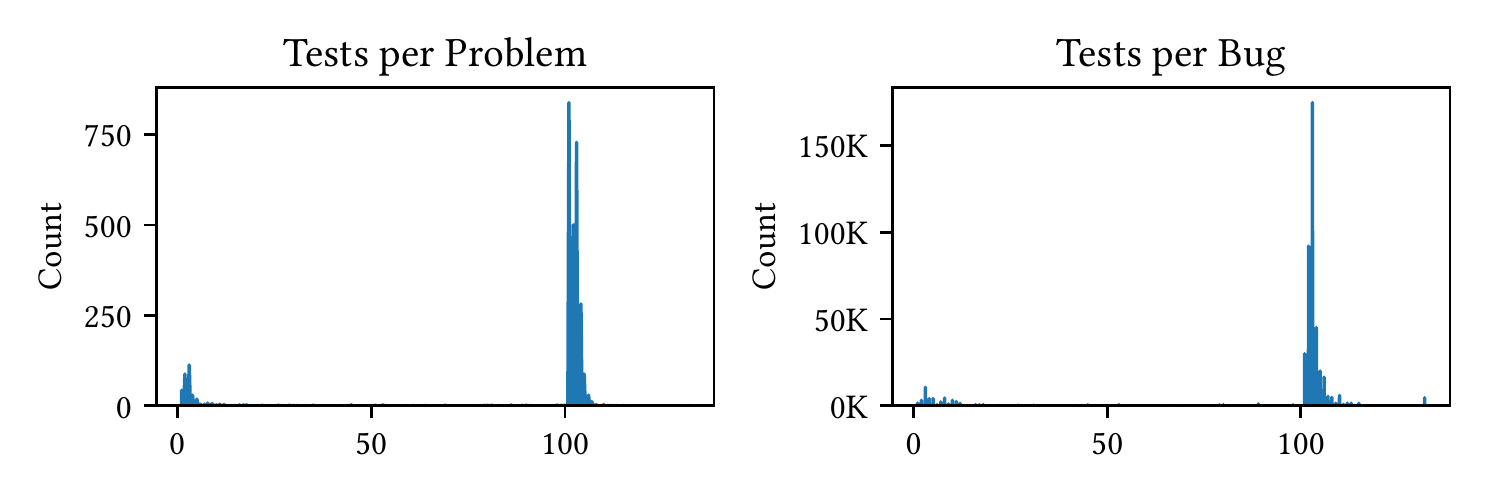}
	\end{center}
	\caption{Distribution of the number of tests per bug and problem in the final dataset.}
	\label{fig:tests-per-problem}
\end{figure}

\begin{figure}[htb]
	\begin{center}
		\includegraphics[width=\linewidth]{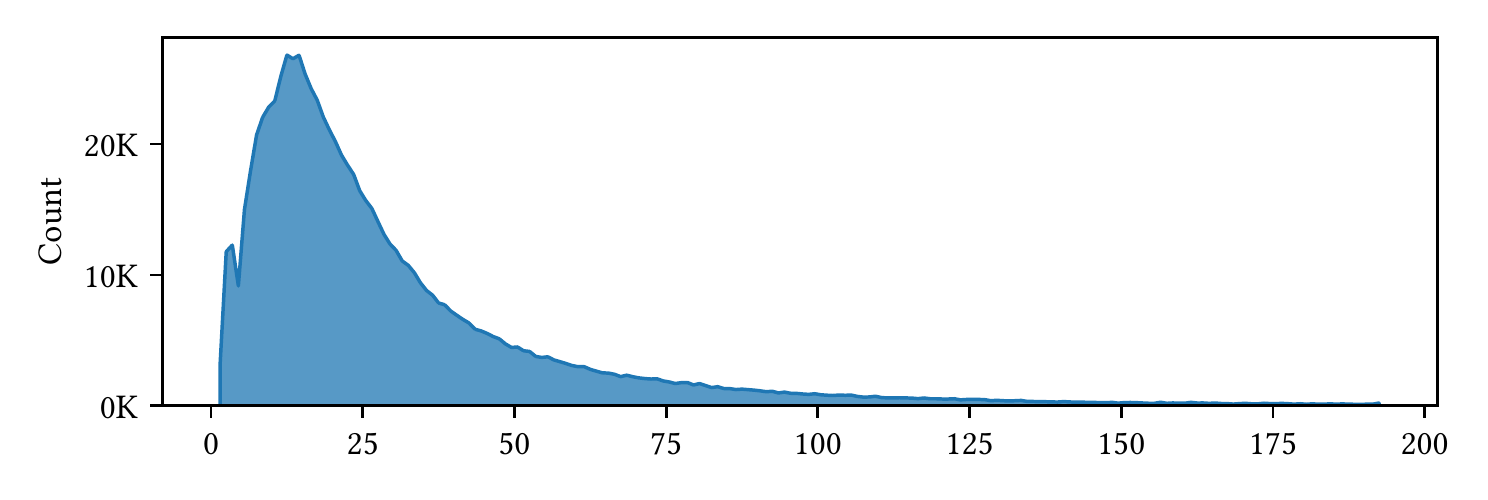}
	\end{center}
	\caption{Distribution of the number of lines of submission programs in the final dataset. For better presentation, the longest 1.3\% of submissions (outliers) are omitted from the plot.}
	\label{fig:locs}
\end{figure}

\lstdefinestyle{trace}{
	basicstyle=\ttfamily,
	language=Python,
	numbers=none,
	moredelim=**[is][{\btHL[fill=gray!30]}]{`}{`},
	escapeinside={(*}{*)},
	numbers=none,
	basicstyle=\fontsize{7pt}{7pt}\selectfont\ttfamily,
	morekeywords={call,line,return,print}
}

\begin{figure}[ht]
    \centering
	\begin{subfigure}{0.4\textwidth}
\begin{lstlisting}[style=trace]
line 1 
line 2 N=3, H=5, W=2
line 3 N=3, H=5, W=2, ans=0
line 4 N=3, H=5, W=2, ans=0, i=0
line 5 N=3, H=5, W=2, ans=0, i=0, x=10, y=3
`line 3 N=3, H=5, W=2, ans=0, i=0, x=10, y=3`
line 4 N=3, H=5, W=2, ans=0, i=1, x=10, y=3
line 5 N=3, H=5, W=2, ans=0, i=1, x=5, y=2
line 6 N=3, H=5, W=2, ans=0, i=1, x=5, y=2
line 3 N=3, H=5, W=2, ans=1, i=1, x=5, y=2
line 4 N=3, H=5, W=2, ans=1, i=2, x=5, y=2
line 5 N=3, H=5, W=2, ans=1, i=2, x=2, y=5
line 3 N=3, H=5, W=2, ans=1, i=2, x=2, y=5
line 7 N=3, H=5, W=2, ans=1, i=2, x=2, y=5
\end{lstlisting}
\caption{buggy execution trace}
\end{subfigure}
\hfill
\begin{subfigure}{0.4\textwidth}
\begin{lstlisting}[style=trace]
line 1 
line 2 N=3, H=5, W=2
line 3 N=3, H=5, W=2, ans=0
line 4 N=3, H=5, W=2, ans=0, i=0
line 5 N=3, H=5, W=2, ans=0, i=0, x=10, y=3
`line 6 N=3, H=5, W=2, ans=0, i=0, x=10, y=3`
line 3 N=3, H=5, W=2, ans=1, i=0, x=10, y=3
line 4 N=3, H=5, W=2, ans=1, i=1, x=10, y=3
line 5 N=3, H=5, W=2, ans=1, i=1, x=5, y=2
line 6 N=3, H=5, W=2, ans=1, i=1, x=5, y=2
line 3 N=3, H=5, W=2, ans=2, i=1, x=5, y=2
line 4 N=3, H=5, W=2, ans=2, i=2, x=5, y=2
line 5 N=3, H=5, W=2, ans=2, i=2, x=2, y=5
line 3 N=3, H=5, W=2, ans=2, i=2, x=2, y=5
line 7 N=3, H=5, W=2, ans=2, i=2, x=2, y=5
\end{lstlisting}
\caption{fixed execution trace}
\end{subfigure}
\begin{subfigure}[b]{\textwidth}
\begin{center}
\begin{tabular}{c}	
\begin{lstlisting}[language=python,style=patch,breaklines=true, showstringspaces=false]
N,H,W=map(int,input().split())
ans=0
for i in range(N):
    x,y=map(int,input().split())
    if x(*{\btHL[fill=red!30]<=}{\btHL[fill=green!30]>=}*)H and y(*{\btHL[fill=red!30]<=}{\btHL[fill=green!30]>=}*)W:
        ans+=1
print(ans)
\end{lstlisting}
\end{tabular}
\end{center}
\caption{patch}
\end{subfigure}
    \caption{Execution traces of bug \#579877 in the test set. Buggy version on the left, fixed version on the right. The bug is caused by a faulty if-condition on line 5 (bottom). After line 5, the two traces diverge (highlighted in gray).}
    \label{fig:execution-traces}
\end{figure}

\begin{figure*}[htbp]
	\begin{minipage}{0.5\textwidth}
		\begin{lstlisting}[language=Ruby,style=patch,numbers=none,xleftmargin=0pt,xrightmargin=0pt,framesep=0pt]
i , max = gets.split.map(&:to_i)
array = gets.split.map(&:to_i).sort
cnt = 0
((*{\btHL[fill=red!30]0}{\btHL[fill=green!30]1}*)..i-1).each do |n|
  if max < array[n] 
    break
  end
  if n == array.length-1 && max > array[n]
    break
  end
  max -= array[n]
    (*{\btHL[fill=red!30]count}{\btHL[fill=green!30]cnt}*) += 1
  end
  puts cnt		
\end{lstlisting}
	\end{minipage}
\hfill
	\begin{minipage}{0.4\textwidth}
		\footnotesize
		
		{\renewcommand{\arraystretch}{1.6}%
			\begin{tabular}{r|l}
				\textbf{ID} & 609453 \\
				\hline
				\textbf{Language} & Ruby  \\
				\hline
				\textbf{Split} & Train \\
				\hline
				\textbf{Labels} & \makecell[l]{ \texttt{literal.number.integer.change},\\
					\texttt{identifier.change}} \\
				\hline
				\thead[r]{\textbf{Failed}\\\textbf{Tests}} & 
				\tiny 
				{\renewcommand{\arraystretch}{1}%
					\begin{tabular}{c|c|c}
						\textbf{Input} & \textbf{\thead{Expected\\Output}} & \textbf{\thead{Actual\\Output}} \\
						... & \texttt{6} & \textit{error} \\
						... & ... & ... \\
				\end{tabular}} \\
				\hline
				\thead[r]{\textbf{Passed}\\\textbf{Tests}} & 
				\tiny
				{\renewcommand{\arraystretch}{1}%
					\begin{tabular}{c|c|c}
						\textbf{Input} & \textbf{\thead{Expected\\Output}} & \textbf{\thead{Actual\\Output}} \\
						... & \texttt{0} & \texttt{0\textbackslash n} \\
						... & ... & ... \\
				\end{tabular}} \\
				\hline
				\thead[r]{\textbf{Error}\\\textbf{Messages}} &
\begin{lstlisting}[basicstyle=\fontsize{5pt}{5pt}\selectfont\ttfamily]
file.rb:12:in `block in <main>':
undefined method `+' for nil:NilClass
(NoMethodError)
	from file.rb:4:in `each'
	from file.rb:4:in `<main>'
\end{lstlisting} \\
		\end{tabular}}
	\end{minipage}
	\caption{Training sample in the Ruby programming language. The submission uses wrong loop bounds and the wrong variable, causing an exception to be thrown. Empty lines in the original were removed and the line numbers in the error message updated accordingly.}
	\label{fig:example-bug-ruby}
\end{figure*}

\begin{figure*}[htbp]
	\begin{minipage}{0.45\textwidth}
		\begin{lstlisting}[language=Java,style=patch,numbers=none,xleftmargin=0pt,xrightmargin=0pt,framesep=0pt]
import java.util.Scanner;
class Main {
  public static void main(String[] args) {
    Scanner sc = new Scanner(System.in);
      int W = sc.nextInt();
      int H = sc.nextInt();
      int x = sc.nextInt();
      int y = sc.nextInt();
      int r = sc.nextInt();
      if ((x (*{\btHL[fill=red!30]+}{\btHL[fill=green!30]-}*) r) < 0 || W < (x + r) ||
          (y (*{\btHL[fill=red!30]+}{\btHL[fill=green!30]-}*) r) < 0 || H < (y + r)) {
         System.out.println("No");
         return;
      }
      System.out.println("Yes");
   }
}
		\end{lstlisting}
	\end{minipage}
	\hfill
	\begin{minipage}{0.45\textwidth}
		\footnotesize
		
		{\renewcommand{\arraystretch}{1.6}%
			\begin{tabular}{r|l}
				\textbf{ID} & 112714 \\
				\hline
				\textbf{Language} & Java  \\
				\hline
				\textbf{Split} & Test \\
				\hline
				\textbf{Labels} & \makecell[l]{
					\tiny\texttt{misc.opposites},\\
					\tiny\texttt{expression.operator.arithmetic.change},\\ \tiny\texttt{control\_flow.branch.if.condition.change}} \\
				\hline
				\thead[r]{\textbf{Failed}\\\textbf{Tests}} & 
				\tiny 
				{\renewcommand{\arraystretch}{1}%
					\begin{tabular}{c|c|c}
						\textbf{Input} & \textbf{\thead{Expected\\Output}} & \textbf{\thead{Actual\\Output}} \\
						... & \texttt{Yes} & \texttt{No} \\
						... & ... & ... \\
				\end{tabular}} \\
				\hline
				\thead[r]{\textbf{Passed}\\\textbf{Tests}} & 
				\tiny
				{\renewcommand{\arraystretch}{1}%
					\begin{tabular}{c|c|c}
						\textbf{Input} & \textbf{\thead{Expected\\Output}} & \textbf{\thead{Actual\\Output}} \\
						... & \texttt{Yes} & \texttt{Yes} \\
						... & ... & ... \\
				\end{tabular}} \\
				\hline
				\thead[r]{\textbf{Error}\\\textbf{Messages}} &
				None \\
		\end{tabular}}
	\end{minipage}
	\caption{Test sample in the Java programming language. The submission contains a buggy if-condition (wrong operator).}
	\label{fig:example-bug-java}
\end{figure*}

\begin{table}[htbp]
    \footnotesize
    \centering
    \caption{Links to candidate evaluation visualization. Users can filter evaluations by label. Each row represents a bug in the given language, each of the 10 columns a candidate. Color represents the evaluation result. Hover a cell to see further details.}
    \label{tab:candidate-vis}

    \begin{tabular}{rl}
         \toprule
         C & \url{https://giganticode.github.io/rbugr/candidate_vis/c.html} \\
         C++ & \url{https://giganticode.github.io/rbugr/candidate_vis/cpp.html} \\
         Java & \url{https://giganticode.github.io/rbugr/candidate_vis/java.html} \\
         Python & \url{https://giganticode.github.io/rbugr/candidate_vis/python.html} \\
         Ruby & \url{https://giganticode.github.io/rbugr/candidate_vis/ruby.html} \\
         Go & \url{https://giganticode.github.io/rbugr/candidate_vis/go.html} \\
         PHP & \url{https://giganticode.github.io/rbugr/candidate_vis/php.html}\\
         JavaScript & \url{https://giganticode.github.io/rbugr/candidate_vis/javascript.html} \\

 \bottomrule
 
    \end{tabular}
\end{table}

\afterpage{%
	\clearpage
\begin{landscape}
\vspace*{\fill}
\begin{figure}[htb]
	\centering
	\includegraphics[width=\linewidth]{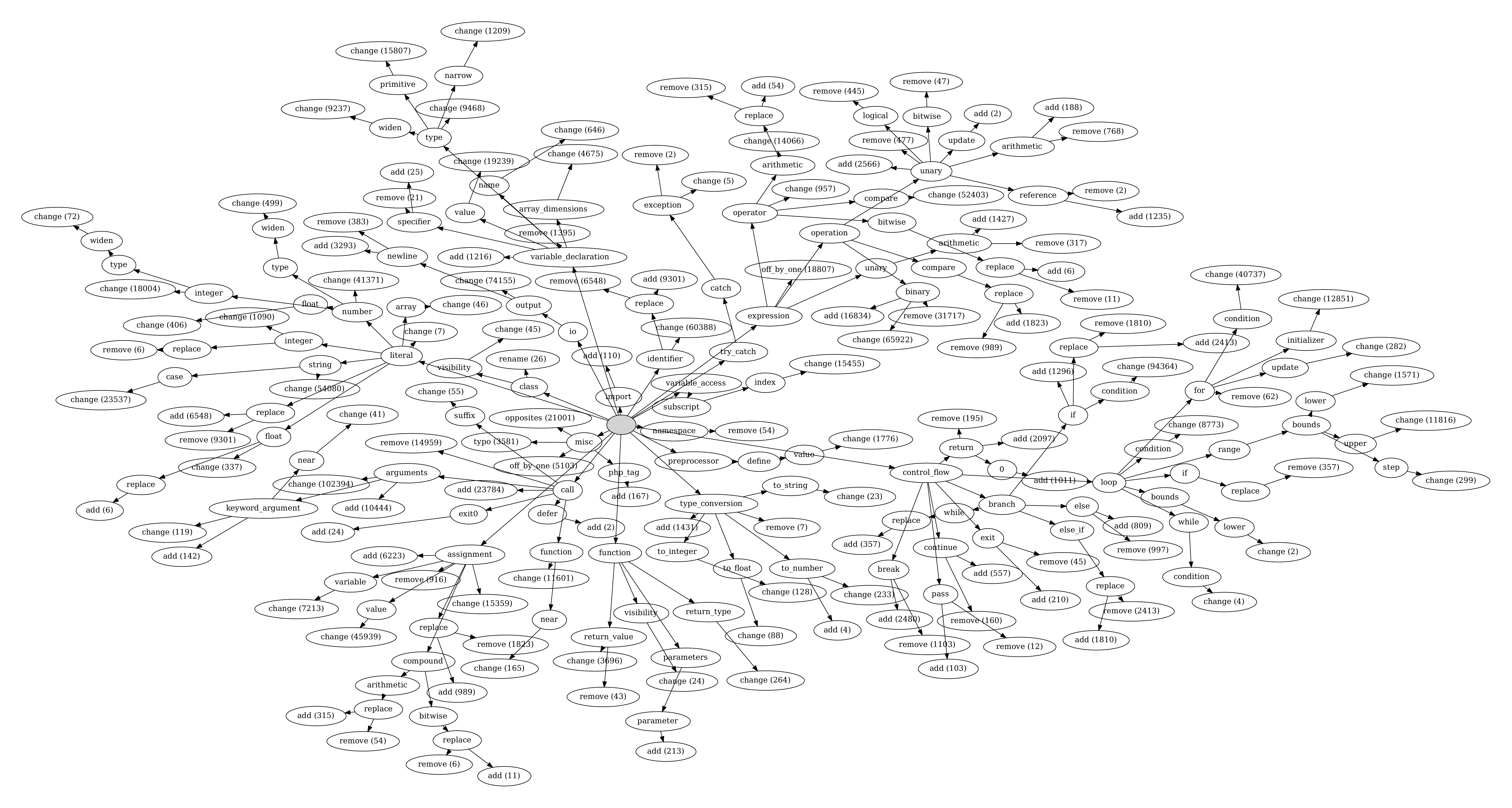}
	\caption{Hierarchical diagram of all labels in the dataset. Leaf nodes represent actual labels; number in parentheses shows label count.}
	\label{fig:labels-diag}
\end{figure}
\vspace*{\fill}
\end{landscape}
\aftergroup}

\end{document}